\documentclass[fleqn,usenatbib]{mnras}

\usepackage{newtxtext,newtxmath}

\usepackage[T1]{fontenc}

\DeclareRobustCommand{\VAN}[3]{#2}
\let\VANthebibliography\thebibliography
\def\thebibliography{\DeclareRobustCommand{\VAN}[3]{##3}\VANthebibliography}

\usepackage{graphicx}	
\usepackage{amsmath}	

\usepackage{amssymb}	

\usepackage{graphicx}	
\usepackage{amsmath}	
\usepackage{amssymb}	

\usepackage{txfonts}
\usepackage{hyperref} 
\usepackage{float} 
\usepackage{natbib}
\usepackage{multirow}
\usepackage{longtable}
\usepackage{numprint}

\newcommand{\abu}{$\log \varepsilon$}

\newcommand{\prione}{Pr\ 221}
\newcommand{\pritwo}{Pr\ 237}

\title[Chemical analysis of two UMP stars]{The Pristine survey XIV: chemical analysis of two ultra-metal-poor stars\thanks{Based on observations collected at the European Southern Observatory under ESO programmes 299.D-5042, 102.D-0766 and 104.B-0305 and on observations made with the Gran Telescopio Canarias (GTC), installed at the Spanish Observatorio del Roque de los Muchachos of the Instituto de Astrofísica de Canarias, in the island of La Palma. Programme ID 39-GTC16/17A.}}

\author[C.~Lardo et al.]{
C.~Lardo,$^{1}$\thanks{E-mail: carmela.lardo2@unibo.it}
L.~Mashonkina,$^{2}$
P.~Jablonka,$^{3,4}$
P.~Bonifacio,$^{4}$
E.~Caffau,$^{4}$
D.~S.~Aguado,$^{5}$
\newauthor
J.~I.~González~Hernández,$^{6,7}$
F.~Sestito,$^{8}$
C.~L.~Kielty,$^{8}$
K.~A.~Venn,$^{8}$
V.~Hill,$^{9}$
E.~Starkenburg,$^{10}$
\newauthor
N.~F.~Martin,$^{11,12}$
T.~Sitnova,$^{2}$
A.~Arentsen,$^{11}$
R.~G.~Carlberg,$^{13}$
J.~F.~Navarro,$^{8}$
and G.~Kordopatis$^{9}$
\\
\\
$^{1}$ Dipartimento di Fisica e Astronomia, Universit\`a degli Studi di Bologna, Via Gobetti 93/2, I-40129 Bologna, Italy\\
$^{1}$ Institute of Astronomy, Russian Academy of Sciences, RU-119017 Moscow, Russia\\
$^{3}$ Laboratoire d'astrophysique, École Polytechnique Fédérale de Lausanne (EPFL), Observatoire, 1290 Versoix, Switzerland\\
$^{4}$ GEPI, Observatoire de Paris, Université PSL, CNRS, 5 Place Jules Janssen, 92190, Meudon, France\\
$^{5}$ Institute of Astronomy, University of Cambridge, Madingley Road, Cambridge CB3 0HA, UK\\
$^{6}$ Instituto de Astrofísica de Canarias, Vía Láctea, 38205 La Laguna, Tenerife, Spain\\
$^{7}$ Universidad de La Laguna, Departamento de Astrofísica, 38206 La Laguna, Tenerife, Spain\\
$^{8}$Department of Physics and Astronomy, University of Victoria, Victoria, BC, V8W 3P2, Canada\\
$^{9}$ Laboratoire Lagrange, Université de Nice Sophia-Antipolis, Observatoire de la Côte d'Azur, France\\
$^{10}$ Kapteyn Astronomical Institute, University of Groningen, Landleven 12, NL-9747 AD Groningen, the Netherlands\\
$^{11}$ Université de Strasbourg, CNRS, Observatoire astronomique de Strasbourg, UMR 7550, F-67000 Strasbourg, France\\
$^{12}$ Max-Planck-Institut für Astronomie, Königstuhl 17, D-69117 Heidelberg, Germany\\
$^{13}$ Department of Astronomy \& Astrophysics, University of Toronto, Toronto, ON M5S 3H4, Canada}

\date{Accepted 2021 September 29. Received 2021 September 29; in original form 2021 June 21}

\pubyear{2020}

\begin{document}
\label{firstpage}
\pagerange{\pageref{firstpage}--\pageref{lastpage}}
\maketitle

\begin{abstract}


Elemental abundances of the most metal-poor stars reflect the conditions in the early Galaxy and the properties of the first stars. We present a spectroscopic follow-up of two ultra metal-poor stars ([Fe/H]<--4.0) identified by the survey {\em Pristine}: Pristine 221.8781+9.7844 and Pristine 237.8588+12.5660 (hereafter~\prione~and~\pritwo, respectively). Combining data with earlier observations, we find a radial velocity of --149.25 $\pm$ 0.27 and --3.18 $\pm$ 0.19 km/s for \prione~ and \pritwo, respectively, with no evidence of variability between 2018 and 2020.

From a one-dimensional (1D) local thermodynamic equilibrium (LTE) analysis, we measure [Fe/H]$_{\rm LTE}$=--4.79 $\pm$ 0.14 for~\prione~and [Fe/H]$_{\rm LTE}$=--4.22 $\pm$ 0.12 for~\pritwo, in good agreement with previous studies. Abundances of Li, Na, Mg, Al, Si, Ca, Ti, Fe, and Sr were derived based on the non-LTE (NLTE) line formation calculations. When NLTE effects are included, we measure slightly higher metallicities: [Fe/H]$_{\rm NLTE}$=--4.40 $\pm$ 0.13 and [Fe/H]$_{\rm NLTE}$=--3.93 $\pm$ 0.12, for~\prione~and~\pritwo, respectively.
Analysis of the G-band yields [C/Fe]$_{\rm 1D-LTE} \leq$ +2.3 and [C/Fe]$_{\rm 1D-LTE} \leq$ +2.0 for~\prione~and~\pritwo. Both stars belong to the low-carbon band. Upper limits on nitrogen abundances are also derived.
Abundances for other elements exhibit good agreement with those of stars with similar parameters.

 
 Finally, to get insight into the properties of their progenitors, we compare NLTE abundances to theoretical yields of zero-metallicity supernovae.
This suggests that the supernovae progenitors had masses ranging from 10.6 to 14.4~M$_{\odot}$ and low-energy explosions with 0.3-1.2 $\times$ 10$^{51}$ erg.
\end{abstract}




\section{Introduction}

The most metal-poor stars in the Galaxy tell us about the conditions found during the first stages of chemical enrichment \citep[e.g.][]{beers05,frebel15}. Given their low metallicities, they are considered to be second-generation objects from the early Universe. Thus, their abundance patterns should reflect the properties of the first massive Population~III (Pop~III) stars.  

Although much effort has been put on the study of this key stellar population over the past 50 years, observations of  the most metal-poor stars populating the Milky Way (MW) and its close satellites are relatively scarce. Indeed, the numbers of known extremely, ultra, and hyper metal-poor stars (stars with [Fe/H] $<$ --3.0, [Fe/H]  $<$ --4.0, and [Fe/H] $<$ --5.0; \citealp{beers05}) are still exceedingly small \citep[e.g., see the compilation presented in][]{sestito19} and fewer than 20 stars are known to have iron abundance below [Fe/H] $<$ --4.5 \citep{ norris07,christlieb04,caffau11,caffau16,hansen14,keller14,allendeprieto15,bonifacio15,frebel05,frebel08,aguado18a,aguado18b,starkenburg18}.

The wide-field photometric survey {\em Pristine} has been especially designed to identify extremely metal-poor stars in an efficient manner \citep{starkenburg17}\footnote{ {\em Pristine} reaches a $\sim$23\% success rate in detecting [Fe/H]$\leq$--3 stars whereas previous surveys have a success rate of $\leq$~5\% \citep{aguado19}.}.
Whereas many surveys for the search of extremely metal-poor stars have relied on spectra --{\em HK Survey} \citep{beers85,beers92}; {\em Hamburg ESO Survey} \citep{christlieb02,christlieb08,frebel05,frebel06}; {\em RAVE} \citep{rave,fulbright10}); {\em Pristine} -- like the {\em SkyMapper Southern Sky} \citep{keller07,wolf18}, the {\em S-PLUS} \citep{splus,whitten21}, and the {\em J-PLUS} \citep{jplus} surveys; uses a photometric filter to isolate specific spectral regions that are sensitive to metallicity. Indeed, the {\em Pristine} \ion{Ca}{ii} HK filter is centred on the lines of ionised calcium in the near ultraviolet which are extremely strong --and thus easily detectable-- even at very low-metallicities.
Medium- and high-resolution spectroscopic follow-ups have been secured to fully characterise candidate metal-poor stars pre-selected by {\em Pristine} and study their chemical composition in great detail \citep{youakim17,starkenburg18,bonifacio19,aguado19,arentsen20a,caffau17,caffau20,venn20,kielty20}.

Pristine\_221.8781+09.7844 (\prione~hereafter) was first highlighted in the {\it Pristine} photometry as an extremely metal-poor star and it was followed up spectroscopically using the medium-resolution Intermediate-dispersion Spectrograph and Imaging System (ISIS) at the 4.2m William Herschel Telescope (La Palma, Spain). This discovery led to the acquisition of high-resolution UVES spectra taken at the Kueyen 8.2m Very Large Telescope (VLT; Paranal, Chile). The analysis of the high-resolution data confirmed the ultra metal-poor nature of the star ([Fe/H] = --4.66 $\pm$ 0.13) with an enhancement of 0.3-0.4 dex in $\alpha$-elements and a remarkably low carbon abundance \citep{starkenburg18}. These authors were able to put an upper limit of $\log \epsilon$(C) $\leq$ 5.6 ([C/Fe] $\leq$ +1.76), a value well below typical values in the ultra metal-poor regime, suggesting that the star is carbon-normal \citep[see discussion in][]{starkenburg18}. This makes~\prione~one of the  three most metal-poor star (all metals included) known to date together with SDSS\ J102915+172927, a star with [Fe/H]$_{\rm LTE}$=--4.73 ([Fe/H]$_{\rm NLTE}$=--4.60) and carbon abundance $\log \epsilon$(C) $\leq$ 4.3 \citep{caffau12} and HE~0557--4840 \citep{norris07,norris12}, a 
red giant branch (RGB) star with [Fe/H] = --4.8 and only a moderate enhancement in the C, N, and O abundances ([C/Fe] = +1.1, [N/Fe] < +0.1, and [O/Fe] = +1.4; all corrected for 3D effects, see \citealp{norris12}).

The ongoing low-resolution campaign to confirm the extremely and ultra metal-poor candidates identified in {\em Pristine} photometry has yielded another interesting object: Pristine\ 237.8588+12.5660 (\pritwo~hereafter). \citet{aguado19}  measured a metallicity of [Fe/H] =--3.88 $\pm$ 0.08 from the \ion{Ca}{ii} K line  with a nearly flat CH-band from the medium-resolution spectrum obtained with the Intermediate Dispersion Spectrograph (IDS) mounted on the 2.5m Isaac Newton Telescope. From a high-resolution analysis, \citet{kielty20} determined a metallicity of [Fe~I/H]$\leq$--4.28 from the analysis of five Fe~I lines. However, due to the limited wavelength coverage of their high-resolution spectra, the only elements whose abundances could be measured were lithium, magnesium, and calcium, whereas other key elements have only upper limits to their abundances.

At extremely low metallicities, stars typically show a somewhat high carbon abundance \citep[e.g.][]{spite13,bonifacio15,norris13,hansen16,yoon18,rasmussen20}. Interestingly, this seems to be a function of the stellar metallicity, placing them all around the absolute C abundance of \abu~(C) $\sim$6.3 $\pm$ 0.5.
This in turn, suggests that {\em (i)} first stars generally produced copious amounts of carbon during late stages of stellar evolution and/or their explosions \citep[e.g.][]{umeda03,meynet06,ekstrom08,choplin18} and {\em (ii)} a certain amount of metals available in the gas cloud out of which second generation objects were formed might be needed to cool down,  fragment and eventually form low-mass stars \citep{frebel07}. The existence of a class of ultra metal-poor stars
which are not severely carbon-enhanced therefore has enormous implications for the formation of low-mass stars at high-redshifts.

In order to get further insight into the physics of early star formation and the properties of the first supernovae, it is important to fully characterise the abundance pattern and, in particular, the carbon content of these interesting objects. In this paper we present and analyse new high-resolution observations of~\prione~and~\pritwo~to complement and expand the studies of~\citet{starkenburg18}, \citet{aguado19}, and~\citet{kielty20}.

The paper is organised as follows. In~\S\ref{observations} we discuss the observations and data reduction. \S\ref{abundance_analysis} contains a description of the stellar parameters and abundance derivations;~\S\ref{results} presents the abundance results. In particular, in~\S\ref{const_on_carbon} we focus on carbon abundances. In~\S\ref{sec:progenitors} we compare elemental abundances to a set of theoretical yields of zero-metallicity supernovae to characterise the properties of the UMP progenitor supernovae. 
Finally, we summarise our findings and draw the conclusions in~\S\ref{conclusions}.


\section{Observations and Data reduction}\label{observations}

Observations were carried out with the UVES spectrograph \citep{dekker10} between 16 Mar to 22 Mar 2020 in delegated visitor mode, with relatively good atmospheric conditions, e.g., the typical seeing was around 1.0\arcsec. 
For these high-resolution observations, we used the standard setting DIC1 390+580, that covers the wavelength intervals 3300-4500\AA~in the blue arm, and 4790-5760~\AA~and 5840-6800~\AA~in the red arm. The adopted set-up was combined with a slit width of 1\arcsec with 1 $\times$ 1 binning on the CCD. Observations were split in 14 and 15 observing blocks, respectively for \prione~and~\pritwo, corresponding to 14:45 hr and 13 hr of total integration time.

UVES Spectra were reduced using the ESO Common Pipeline Library, UVES pipeline version 5.10.4\footnote{https://www.eso.org/sci/software/pipelines/uves/uves-pipe-recipes.html}. The pipeline processing steps include bias subtraction, background subtraction optimal extraction, automatic sky subtraction, flat-fielding of the extracted spectra,  wavelength calibration constructed from the Th-Ar lamp exposure, resampling at a constant wavelength step and optimal merging of the echelle orders. 

\begin{figure}
\includegraphics[width=0.90\columnwidth]{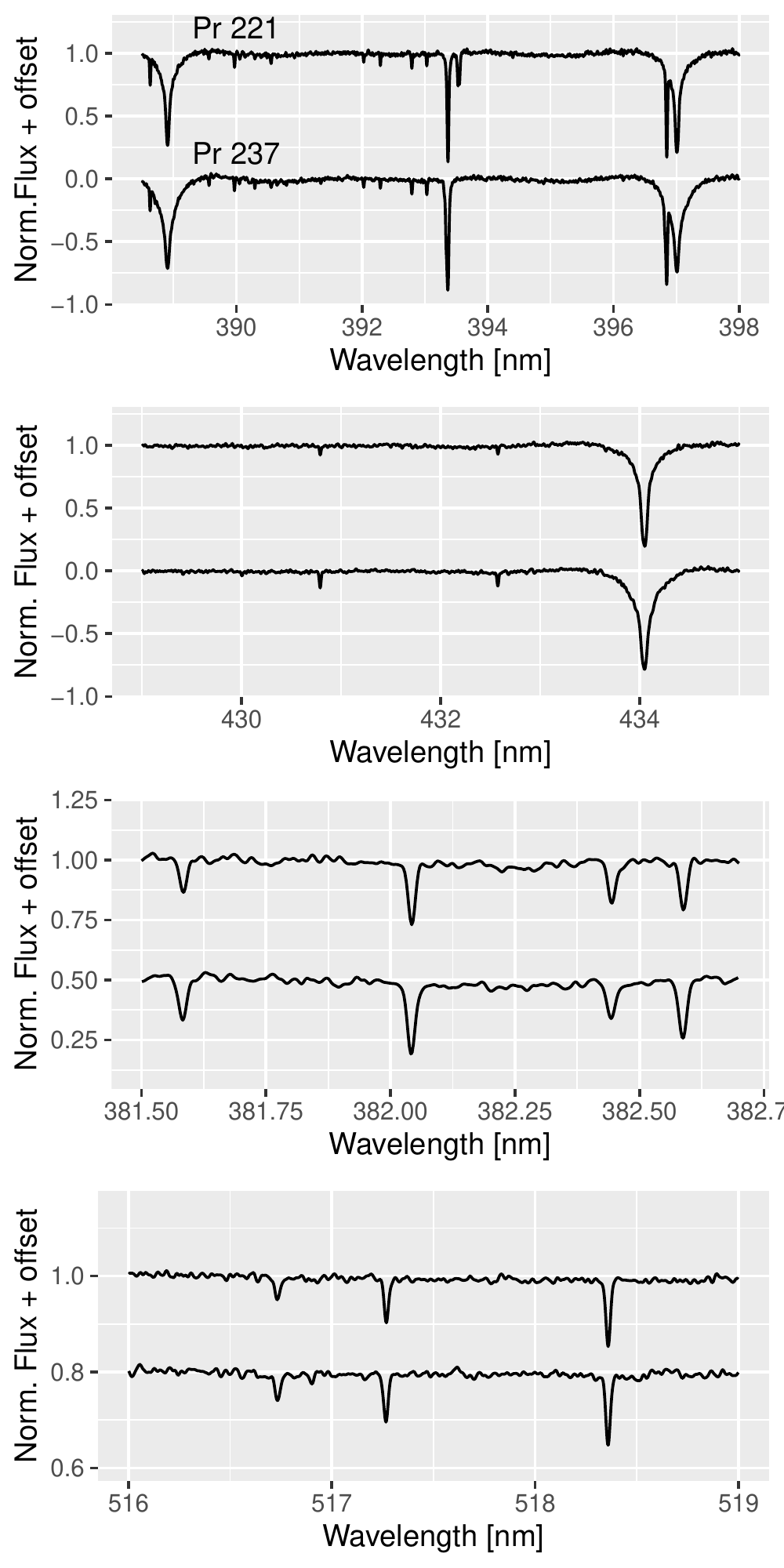}
      \caption{The figures shows (from top to bottom) a detail of the UVES spectra of \prione~(upper spectrum)~and~\pritwo~(lower spectrum)~around the \ion{Ca}{ii} HK, the CH-band, a zoom in the vicinity of the strong iron lines around 3820~\AA, and the green \ion{Mg}{i} b triplet.}
\label{fig:raw_spectra}
\end{figure}

The reduced spectra were then corrected for the barycentric velocity.
Spectra shifted to the rest wavelength and smoothed to the resolution of 30\ 000 are combined as final step. Observations for~\prione~were complemented with data from the DDT UVES programme (Programme 299.D-5042; which was using the same observational setup; see \citealp{starkenburg18} for details). The approximate signal-to-noise ratio (SNR) per pixel of the final combined spectra are 90 at 4000~\AA~, 160 a 5300~\AA, and 300 at 6700~\AA. In Figure~\ref{fig:raw_spectra} we show some selected spectral intervals of the analysed UVES spectra to illustrate the high-quality of the data.

\subsection{Medium-resolution data}

\pritwo~was also observed using the medium-resolution Optical System for Imaging and low-Intermediate-Resolution Integrated Spectroscopy (OSIRIS) spectrograph at the 10.4m  Gran Telescopio Canarias (GTC) located in the Roque de los Muchachos observatory (La Palma, Spain). We dedicated 1hr of GTC observing time to obtain a long-slit spectrum with the R2500U VPH grating covering the spectral range 3440-4610~\AA~and 1.0\arcsec slit, leading to a resolving power of R$\sim$2400. OSIRIS spectra were reduced in a standard manner including bias, flat-field correction, extraction, sky subtraction, and wavelength calibration. The mean SNR of the combined OSIRIS spectrum of \pritwo~is 170.

\subsection{Radial velocity measurements}

Radial velocities (RVs) have been derived using {\sc daospec}~\citep{daospec} on individual UVES exposures. The blue and red arms were analysed independently and the agreement between the two radial velocity estimates is good (i.e. the difference is less than 1.5 km s$^{-1}$ for both stars). The radial heliocentric velocity of the final spectrum is measured to be --149.25 $\pm$ 0.27 km s$^{\rm{-1}}$ ($\sigma$=1.1 km s$^{\rm{-1}}$)  and  --3.18 $\pm$ 0.19 km s$^{\rm{-1}}$ ($\sigma$=0.7 km s$^{\rm{-1}}$)  for~\prione~and~\pritwo, respectively. 
These derived values are in excellent agreement with those derived by \citet{starkenburg18} for~\prione~(--149.0 $\pm$ 0.5 km s$^{\rm{-1}}$) and \citet{kielty20} for~\pritwo~(--3.8 $\pm$ 0.9 km s$^{\rm{-1}}$). 

For both stars, no significant velocity variations are measured on different exposures. Therefore, we have no evidence that these stars might be influenced by any binary companion and we continue with the assumption that the stars are single for the remainder of the paper.
A complete discussion of the orbital properties in the Galaxy of the target stars can be found in \citet{sestito19,sestito20} and \citet{kielty20}.

RV measurements for the two stars analysed in this paper are listed in Table~\ref{table:parameters}, along with other useful information.

\begin{table}[!h]
\caption{Position, photometry from SDSS DR13 \citep{albareti17} and adopted stellar parameters for the two stars analysed in this study.}
\label{table:parameters}    
\centering                        
\begin{tabular}{l r r}       
	 \cline{2-3}
	  \cline{2-3}
	 
                        &Pr\ 221.8781+9.7844    &	 Pr\ 237.8588+12.566 \\
  \hline	
  RA  (J2000)		    &    14:47:30.7         &	 15:51:26.1    \\      
  Dec (J2000)		    &    +09:47:03.70       &	 +12:33:57.60  \\     
  CaHK$_0$ 			    &     16.87 	        &	 15.84 \\ 	      
  g$_0$     			&     16.51   	        &	 15.58 \\   	      
  i$_0$     			&     16.04 	        &	 15.30 \\ 	      
  RV	[km s$^{-1}$]	&     --149.25$\pm$1.1 &	--3.18$\pm$0.37 \\     
  $\log$(g) [dex]		&     3.03$\pm$0.10      &	 3.61$\pm$0.10    \\     
  T$_{\rm eff}$ [K]		&     5683$\pm$134      &	 6148$\pm$175  \\     

\hline                                  
\end{tabular}
\end{table}


\section{Chemical Analysis}\label{abundance_analysis}

\subsection{Stellar Parameters}\label{parameters}

We derive stellar parameters following the approach outlined in \citet{kielty20}.
Briefly, effective temperature (T$_{\rm eff}$) is computed using the \citet{mucciarelli20} colour-T$_{\rm eff}$ relation, which combines effective temperatures derived by \citet{gonzalez09} through the infrared flux method (IRFM) with the photometry available from the second release of the ESA/Gaia mission \citep{gaia16,gaia18} and the K-band magnitudes from the 2MASS database \citep{2mass}. The adopted extinction coefficient from \citet{schlafly11} are re-adapted for Gaia DR2 photometry\footnote{Gaia DR2 extinction relations are from \url{http://stev.oapd.inaf.it/cgi-bin/cmd}.}.
Surface gravity ($\log$g) was determined with the Stefan-Boltzmann equation and the bolometric correction for Gaia filters \citep{andrae18}. The stellar mass is treated as a flat prior, spanning a range from 0.5 to 1 M$_{\odot}$. Parallaxes are from Gaia DR2 \citep{gaia16,gaia18}.

The \citet{mucciarelli20} temperature calibration requires assumption about whether the star is a dwarf or giant and a value for the metallicity. In our computation we assume a metallicity [Fe/H]=--4.0 $\pm$ 0.3 and use MIST/MESA isochrones \citep{paxton11,dotter16} to preliminary select between the dwarf/giant solution. Then we iterate the inference of the \citet{mucciarelli20} temperatures using the $\log$g from Stefan-Boltzmann as input, and then again the $\log$g  with the new temperature. We did these iterations 1000 times to check for convergence or changes under small perturbations. 
Uncertainties associated to the atmospheric parameters are derived through Monte Carlo sampling of the input data taking into account the uncertainties associated to photometry, metallicity, distance, and extinction \citep[e.g.][]{kielty20}.

For~\prione, we found T$_{\rm eff}$=  5683 $\pm$ 134 K and $\log$g= 3.03 $\pm$ 0.10. The derived stellar parameters are  consistent with those derived in the studies of \citet{sestito19} (T$_{\rm eff}$=5710$\pm$65 K, $\log$g=3.1$\pm$0.1), and \citet{starkenburg18} (T$_{\rm eff}$=5792$\pm$100 K, e $\log$g=3.5$\pm$0.5), who adopted a slightly different approach to estimate atmospheric parameters. In \citet{starkenburg18}, an effective temperature of T$_{\rm eff}^{\rm IRFM}$= 5877 $\pm$ 62~K was derived using the IRFM of \citet{gonzalez09} and a surface gravity of  $\log$g=3.5. Since the available photometry from 2MASS \citep{2mass}
has too large uncertainties to be used to infer reliable temperatures, these derivations rely on the infrared UKIDSS JHK magnitudes \citep{lawrence07} transformed into the 2MASS system and $V$  = 16.356 $\pm$ 0.020 from \citet{apass}.
Magnitudes were dereddened  E(B--V)= 0.024 from \citet{schlegel98}.
The same method provides for each of the infrared UKIDSS $JHK$ magnitudes \citep{lawrence07} transformed into the 2MASS system very consistent values of T$_{\rm eff}^{\rm IRFM}$ = 5883 $\pm$ 77~K,  5878 $\pm$ 54~K, and  5871 $\pm$ 60~K; for the $J$, $H$,and $K_s$ band; respectively. Changing the surface gravity value from $\log$g= 3.5 to 3.0 (as assumed here) would yield an effective temperature only 13~K hotter (T$_{\rm eff}^{\rm IRFM}$= 5884 $\pm$ 62~K). 

For~\pritwo~we adopted T$_{\rm eff}$= 6148 $\pm$ 175 K and $\log$g= 3.61 $\pm$ 0.10 as in \citet{kielty20}.
The infrared 2MASS magnitudes $J$  = 14.509 $\pm$ 0.035, $H$ = 14.266 $\pm$ 0.042, and 
$K_s$ = 14.257 $\pm$ 0.060 are also in this case not accurate enough to infer reliable temperatures. Unfortunately, no UKIDSS $JHK$ infrared magnitudes are available for ~\pritwo.  Using the 2MASS infrared magnitudes and assuming $V$ = 15.611 $\pm$ 0.022 \citep{apass} and reddening E(B-V) = 0.039 \citep{schlegel98}, we derive for~\pritwo~ T$_{\rm eff}^{\rm IRFM}$= 6330 $\pm$ 200~K \citep{gonzalez09}.

The effective temperatures derived from the Gaia photometry differ by $\Delta$~T$_{\rm eff} \sim$ --180 ~K from those derived using ground based photometry.  With the available spectra, an inspection of the wings of the Balmer lines does not favour one solution over the other. The IRFM method requires accurate infrared photometry and reddening to estimate reliable temperatures. On the other hand, the \citet{mucciarelli20} calibration rely on all-sky, superb quality photometry from Gaia. Thus, we decided to use the \citet{mucciarelli20} calibration to estimate effective temperature and adopt the \citet{kielty20} procedure to derive stellar parameters. In \S~\ref{sec:abundances} we discuss how the derived abundances would change because of the uncertainties on the atmospheric parameters.

\begin{table*}

\centering 
\npdecimalsign{.}
\nprounddigits{2}
\footnotesize
\caption{Equivalent widths and associated LTE abundances for~\prione~and~\pritwo. The corresponding NLTE abundances are also listed (see text). Data for the atomic transitions analysed are also reported.}             
\label{table:PR221EW}      
\begin{tabular}{r n{4}{2} n{2}{2} n{2}{2} n{3}{2} n{2}{2}  n{2}{2}  n{2}{2} n{2}{2} n{2}{2} n{2}{2} n{2}{2}} 
\hline\hline
 \multicolumn{4}{c}{}     & \multicolumn{4}{c}{Pristine\_221.8781+09.7844} & \multicolumn{4}{c}{Pristine\_237.8588+12.5660}\\ 
 {Ion}    &  {$\lambda$}  & {$\log$(gf)} & {E$_{\rm exc}$}  &  {EW} & {$\sigma$ {\rm EW}} & \multicolumn{2}{c}{\abu (X)} & {EW} & { $\sigma$ {\rm EW}} &  \multicolumn{2}{c}{\abu (X)}\\
\cline{7-8}
\cline{11-12}
 {} &   {(\AA)}      &    {}         &   {(eV)}   &  {(m\AA)} & {(m\AA)} &  {LTE}    &  {NLTE}  &  {(m\AA)} & {(m\AA)}& {LTE}    & {NLTE}\\
\hline
\ion{Li}{i}    &   6707.768   &   0.17     &   0.0	 &   9.4	  &   0.45	   &   1.327    &  1.307    &  10.6	   &      0.55  	&  1.737    &  1.717	\\
\ion{Na}{i}    &   5889.958   &   0.108    &   0.108     &   13.7	  &   0.48	   &   1.841    &  1.826    &     	   &       		&	    &    	\\
\ion{Mg}{i}    &   3829.355   &   -0.227   &   2.71      &   24.0	  &   0.85	   &   3.135    &  3.482    &  20.0	   &      1.38  	&  3.342    &  3.629	\\
 	       &   3832.304   &   0.125    &   2.71      &   50.2	  &   1.87	   &   3.292    &  3.649    &     	   &    		&	    &     	\\
               &   5172.684   &   -0.45    &   2.71      &	25.2	  &   0.86         &   3.301	&  3.611    &  27.7	   &      0.99  	&  3.673    &  3.940	\\
               &   5183.604   &   -0.24    &   2.72      &	39.2	  &   1.45         &   3.372	&  3.659    &  41.2	   &      1.34  	&  3.726    &  3.972	\\
\ion{Al}{i}    &   3961.521   &   -0.333   &   0.01      &   8.9	  &   0.26	   &   1.382    &  2.06     &  6.8	   &      0.79  	&  1.645    &  2.245	\\
\ion{Si}{i}    &   3905.523   &   -1.041   &   1.91      &   13.5	  &   1.06	   &   2.852    &  3.18     &     	   &       		&     	    &     	\\
\ion{Ca}{i}    &   4226.728   &   0.244    &   0.0	 &   35.0	  &   1.58	   &   1.773    &  2.073    &  28.2	   &      1.61  	&  2.047    &  2.311	\\
\ion{Ca}{ii}   &   3933.66    &   0.111    &   0.111     &  	          &   	           &   2.218    &  2.174    &     	   &       		&    	    &     	\\
\ion{Sc}{ii}   &   4246.822   &   0.242    &   0.310     &   6.60         &   0.21         &  -1.099    &           &              &       		&    	    &           \\
\ion{Ti}{ii}   &   3759.291   &   0.28     &   0.61      &   28.8	  &   0.44	   &   0.385    &  0.53     &  46.3	   &      0.72  	&  1.168    &  1.288 	\\
               &   3761.321   &   0.18     &   0.57      &	33.5	  &   1.51         &   0.554	&  0.69     &  46.4	   &      0.63  	&  1.239    &  1.349 	\\
               &   3913.461   &   -0.36    &   1.12      &	          &      	   &    	&           &  7.1	   &	 0.34	        &  1.154    &  1.214 	\\
               &   4300.043   &   -0.46    &   1.18      &	          &      	   &    	&           &  8.1	   &	 0.4	        &  1.34     &  1.490 	\\
               &   4395.031   &   -0.54    &   1.08      &	          &      	   &    	&           &  7.6	   &	 0.54	        &  1.291    &  1.401 	\\
\ion{Cr}{i}    &   4254.336   &  -0.108    &   0.000     &                &                &            &           &  5.70        &               &  <1.482   & 	   	\\
\ion{Fe}{i}    &   3758.233   &   -0.027   &   0.96	 &   32.3	  &   0.56	   &   2.529	&  2.922    &  36.1	   &	  1.44  	&  3.088    &  3.363	\\
               &   3763.789   &   -0.238   &   0.99	 &   25.5	  &   1.08	   &   2.613	&  3.0      &  29.8	   &	  1.23  	&  3.188    &  3.467	\\
               &   3767.192   &   -0.389   &   1.01	 &		  &		   &		&	    &  21.1	   &	  0.5		&  3.142    &  3.420	\\
               &   3787.88    &   -0.859   &   1.01	 &		  &		   &		&	    &  18.2	   &	  1.36  	&  3.525    &  3.795	\\
               &   3815.84    &   0.232    &   1.48	 &   23.2	  &   0.23	   &   2.578	&  3.054    &  30.7	   &	  0.77  	&  3.195    &  3.553	\\
               &   3820.425   &   0.119    &   0.86	 &   48.5	  &   0.63	   &   2.624	&  3.023    &		   &			&	    &		\\
               &   3824.443   &   -1.362   &   0.0	 &   32.2	  &   0.48	   &   2.876	&  3.243    &  29.1	   &	  1.83  	&  3.354    &  3.629  \\
               &   3825.881   &   -0.037   &   0.91	 &   37.0	  &   0.31	   &   2.587	&  2.986    &  46.0	   &	  0.94  	&  3.261    &  3.537  \\
               &   3827.822   &   0.062    &   1.56	 &   15.2	  &   0.4	   &   2.581	&  3.033    &  19.3	   &	  0.43  	&  3.143    &  3.484  \\
               &   3840.437   &   -0.506   &   0.99	 &   14.5	  &   0.48	   &   2.55	&  2.932    &  17.4	   &	  0.81  	&  3.121    &  3.399  \\
               &   3849.966   &   -0.871   &   1.01	 &   10.7	  &   1.09	   &   2.78	&  3.153    &  15.6	   &	  0.71  	&  3.446    &  3.719  \\
               &   3856.371   &   -1.286   &   0.05	 &   30.7	  &   0.52	   &   2.815	&  3.183    &  25.2	   &	  2.16  	&  3.23     &  3.505  \\
               &   3859.911   &   -0.71    &   0.0	 &   60.6	  &   1.02	   &   2.869	&  3.242    &  59.3	   &	  0.78  	&  3.384    &  3.654  \\
               &   3865.523   &   -0.982   &   1.01	 &   10.3	  &   1.04	   &   2.871	&  3.243    &  7.5	   &	  0.38  	&  3.189    &  3.461  \\
               &   3878.018   &   -0.914   &   0.96	 &   7.5	  &   0.7	   &   2.594	&  2.966    &  10.2	   &	  0.37  	&  3.219    &  3.491  \\
               &   3878.573   &   -1.379   &   0.09	 &   27.1	  &   0.96	   &   2.858	&  3.224    &  22.1	   &	  0.78  	&  3.274    &  3.548  \\
               &   3886.282   &   -1.076   &   0.05	 &   49.7	  &   2.74	   &   3.014	&  3.387    &		   &			&	    &		\\
               &   3895.656   &   -1.67    &   0.11	 &		  &		   &		&	    &  8.4	   &	  0.92  	&  3.078    &   3.349	 \\
               &   3899.707   &   -1.531   &   0.09	 &   19.1	  &   0.61	   &   2.798	&  3.162    &  17.0	   &	  0.52  	&  3.275    &   3.548	 \\
               &   3902.945   &   -0.466   &   1.56	 &   8.3	  &   0.44	   &   2.797	&  3.21     &  14.0	   &	  0.43  	&  3.489    &   3.795	 \\
               &   3920.257   &   -1.746   &   0.12	 &   12.9	  &   0.56	   &   2.836	&  3.198    &  12.6	   &	  0.47  	&  3.363    &   3.634	 \\
               &   3922.911   &   -1.651   &   0.05	 &   17.1	  &   0.31	   &   2.818	&  3.181    &  13.9	   &	  0.42  	&  3.253    &   3.525	 \\
               &   4005.241   &   -0.61    &   1.56	 &		  &		   &		&	    &  10.2	   &	  0.54  	&  3.462    &   3.742	 \\
               &   4045.812   &   0.28     &   1.48	 &   28.3	  &   0.47	   &   2.627	&  3.058    &  40.3	   &	  0.55  	&  3.329    &   3.644	 \\
               &   4063.594   &   0.062    &   1.56	 &   16.7	  &   0.23	   &   2.606	&  3.014    &  27.9	   &	  1.17  	&  3.34     &   3.641	 \\
               &   4071.738   &   -0.022   &   1.61	 &   14.4	  &   0.21	   &   2.662	&  3.064    &  21.2	   &	  0.44  	&  3.303    &   3.600	 \\
               &   4143.868   &   -0.511   &   1.56	 &		  &		   &		&	    &  10.1	   &	  0.58  	&  3.346    &   3.628	 \\
               &   4202.029   &   -0.708   &   1.48	 &		  &		   &		&	    &  8.2	   &	  0.73  	&  3.369    &   3.662	 \\
               &   4260.474   &   0.077    &   2.4	 &		  &		   &		&	    &  8.1	   &	  0.46  	&  3.425    &   3.723	 \\
               &   4271.76    &   -0.164   &   1.48	 &   14.6	  &   0.75	   &   2.667	&  3.086    &  22.6	   &	  0.77  	&  3.349    &   3.656	 \\
               &   4325.762   &   0.006    &   1.61	 &   14.1	  &   0.33	   &   2.599	&  3.032    &  23.0	   &	  0.77  	&  3.3      &   3.616	 \\
               &   4383.544   &   0.2	   &   1.48	 &   29.1	  &   0.21	   &   2.691	&  3.131    &  35.3	   &	  0.17  	&  3.27     &   3.591	 \\
               &   4404.75    &   -0.142   &   1.56	 &   19.1	  &   0.91	   &   2.851	&  3.262    &  22.3	   &	  0.59  	&  3.376    &   3.680	 \\
               &   4415.122   &   -0.615   &   1.61	 &		  &		   &		&	    &  9.5	   &	  0.34  	&  3.446    &   3.738	 \\
               &   5269.537   &   -1.321   &   0.86	 &   11.1	  &   0.51	   &   2.974	&  3.351    &  10.0	   &	  0.55  	&  3.415    &   3.696	 \\
\ion{Ni}{i}    &   3858.297   &   -0.865   &  0.420      & 5.50           &  0.61          &   1.537    &           &  5.50	   &	  0.31          &  2.038    &  	        \\		    
\ion{Sr}{ii}   &   4077.709   &   0.167    &   0.0	 &   6.0	  &   0.57	   &   -2.379   &  -2.2     &  23.5	   &      1.61  	&  -1.216   &  -1.006	\\
               &   4215.519   &   -0.145   &   0.0	 &      	  &      	   &            &     	    &  12.7	   &      0.66  	&  -1.259   &  -1.089	\\
\ion{Y}{ii}    &   3774.331   &   0.210    &  0.130      & 5.90	          &             & <-1.412 	&           &              &       	        &	    &	   	\\
\ion{Ba}{ii}   &   4934.07   &   0.150   & 0.000	 &          &         & <-2.80 	&   	    &	   	   &      		&	  <-2.10    &	   	\\	  

\ion{Eu}{ii}   &   4129.707   &   0.220    & 0.000	 & 5.90	          &           & <-1.589 	&   	    &	   	   &       		&	    &	   	\\	  
 \hline                                   
\end{tabular}
\end{table*}

\subsection{Abundance measurements}\label{sec:abundances}

\subsubsection{High-resolution UVES data}

We determined the chemical abundances for all the measured elements
(except carbon and nitrogen) through the measurement of the equivalent widths (EWs) of atomic transition lines. 

The atomic line list largely overlaps the one described in \citet{starkenburg18}, except for the atomic data for the Mg~I lines at 5172~\AA~and 5183~\AA, for which we used the experimental $\log$(gf)-values from \citet{aldenius07} and the predicted $\log$(gf)-values from \citet{yan95} for the~\ion{Li}{i} transition at 6707~\AA. 
Equivalent widths were measured with the code {\sc daospec}
\citep{daospec} through the wrapper {\sc 4dao} \citep{4dao,4dao_1}.
Uncertainties on the equivalent width measurements are estimated by {\sc daospec} as the standard deviation of the local flux residuals and represent a 68\% confidence interval of the derived EW.  

Elemental abundances from EWs were measured using the package {\sc Gala} \citep{gala} based on the {\sc width9} code by Kurucz. Model atmospheres were calculated with the {\sc atlas9} code assuming local thermodynamic equilibrium (LTE) and one-dimensional, plane-parallel geometry; starting from the grid of models available on F. Castelli’s website \citep{atlas}. The {\sc atlas9} models employed were computed with the new set of opacity distribution functions and exclude approximate overshooting in calculating the convective flux \citep{atlas}.
We run {\sc Gala} keeping T$_{\rm eff}$, $\log$(g) and microturbulence of the model fixed, allowing its metallicity to vary in order to match the Fe abundance measured from EWs. For all the models we adopted as input metallicity the iron abundances derived by \citet{starkenburg18} and \citet{kielty20}. Only lines with a reduced EW (EWr = $\log$ (EW/$\lambda$)) between --5.9 (corresponding to $\sim$5.5~m\AA~for a line at 4300\AA) and --4.7  (corresponding to $\sim$87~m\AA~for a line at 4300\AA) were considered in the abundance analysis, in order to avoid weak and noisy lines as well as  strong features in the flat part of the curve of growth.
All the lines with equivalent width uncertainties larger than 15\%\footnote{An error of $\pm$15\% on EW measurements typically translates into an abundance error of $\sim$0.1 dex for a line located in the linear part of the curve of growth.} are also excluded. Finally, to compute the abundance of iron, we kept only lines within 3$\sigma$ from the median iron value. In the analysis for this work the microturbulent velocity is fixed at 1.5 km s$^{-1}$. The solar values are all from \citet{lodders09}. For carbon and iron we use the solar abundances of \citet{caffau11}. 

Equivalent width measurements for individual lines and corresponding elemental abundances are reported in Table~\ref{table:PR221EW}.
This table also lists the wavelength, excitation potential (E$_{\rm exc}$) and oscillator strengths ($\log$ (gf)) of the considered atomic transitions. 
The averaged results for each element are listed in Table~\ref{table:abundances}. 
For the elements for which we measured more than one line, we provide in Table~\ref{table:abundances} the dispersion of
the single line measurements around the mean normalised to the root square of the number of used lines 
that can be used to estimate the uncertainties associated with the measurements.

When the line-to-line scatter is relatively small ($\leq$0.1 dex), the error is dominated by the errors due to the uncertainties in the stellar parameters. The error is dominated by the uncertainties associated with the derivation of atmospheric parameters, also in the case where only one line is measured for a given element.
Uncertainties due to the atmospheric parameters are computed varying  one parameter at a time by the corresponding error, while keeping the others fixed. They are listed in  Table~\ref{table:errors} for the two stars.
All elements show a significant change with a $\sim$ 150~K variation in temperature. On the contrary, changes in the adopted microturbulent velocity result in very small abundance variations for all elements. Finally, only abundances for those elements which appear in their ionised states (i.e., \ion{Ca}{ii}, \ion{Ti}{ii}, \ion{Sr}{ii}) are sensitive to changes in surface gravity.  

The recommended error estimates associated to the measured abundances (e[X/H]) are summarised in Table~\ref{table:abundances}.

 \begin{table*}
\caption{Chemical abundances for \prione~and~\pritwo}             
\label{table:abundances}    
\centering       
\footnotesize
\setlength{\tabcolsep}{1pt}
\begin{tabular}{r c r r r c r r r c r r r r r c r r r c r r}       
\hline\hline  
      &  & \multicolumn{10}{c}{Pristine\_221.8781+09.7844} & \multicolumn{10}{c}{Pristine\_237.8588+12.5660}\\ 
      &   & \multicolumn{4}{c}{LTE} &   \multicolumn{4}{c}{NLTE} &  &  & \multicolumn{4}{c}{LTE} &   \multicolumn{4}{c}{NLTE} &  & \\
  \cline{3-10}
  \cline{13-20}  

Ion & \abu (X)$_\odot$ & \abu (X)  & [X/H] & [X/Fe] & $\sigma$ &   \abu (X) &   [X/H] & [X/Fe] & $\sigma$ & e[X/H]& N$_{\rm lin}$ & \abu (X)  & [X/H] & [X/Fe] & $\sigma$ &   \abu (X) &   [X/H] & [X/Fe] & $\sigma$ &  e[X/H]& N$_{\rm lin}$ \\
\hline
\ion{Li}{i}  & 1.10 &     1.33   &  0.23   &  5.02   &	      &    1.31 &  0.21   &  4.61  &	    & 0.20 &  1   & 1.74    &    0.64   &  4.86   &	  &   1.72 &	 0.62  &  4.55   &	  & 0.20 & 1 \\
\ion{CH}{}   & 8.50 &    <6.00   & <--2.50 &  <2.29 &        &         &	  &	   &	    & 	   &  1   & <6.30   &   <--2.20 & < 2.02  &	  &	   &	       &	 &	  &      & 1 \\
\ion{NH}{}   & 7.86 &    <5.00   & <--2.86 &  < 1.93 &        &         &	  &	   &	    &	   &  1   & <6.00   &   <--1.86 & <2.36  &	  &	   &	       &	 &	  &      & 1 \\
\ion{Na}{i}  & 6.30 &     1.84   & --4.46  &  0.33   &	      &    1.83 & --4.47  & --0.07 &	    & 0.10 &  1   &	    &	        &         &	  &	   &	       &	 &	  &      &   \\
\ion{Mg}{i}  & 7.54 &     3.28   & --4.26  &  0.53   & 0.10   &    3.60 & --3.94  &  0.46  & 0.08   & 0.10 &  4   & 3.58    &   --3.96  &  0.26   & 0.21  &   3.85 &  --3.69   &   0.24  & 0.19   & 0.13 & 3 \\
\ion{Al}{i}  & 6.47 &	  1.38   & --5.09  & --0.30  &	      &    2.06 & --4.41  & --0.01 &	    & 0.11 &  1   & 1.64    &   --4.83  & --0.61  &	  &   2.24 &  --4.23   &  --0.30 &	  & 0.15 & 1 \\
\ion{Si}{i}  & 7.52 &	  2.85   & --4.67  &  0.12   &	      &    3.18 & --4.34  &  0.06  &	    & 0.11 &  1   &	    &	        &         &	  &	   &	       &	 &	  &      &   \\
\ion{Ca}{i}  & 6.33 & 	  1.77   & --4.56  &  0.23   &	      &    2.07 & --4.26  &  0.14  &	    & 0.13 &  1   &  2.05   &   --4.28  & --0.06  &	  &   2.31 &  --4.02   &  --0.09 &	  & 0.16 & 1 \\
\ion{Ca}{ii} & 6.33 & 	  2.22   & --4.11  &  0.68   &	      &    2.17 & --4.16  &  0.24  &	    & 0.14 &  1   &	    &	        &         &	  &	   &	       &	 &	  &      &   \\
\ion{Sc}{ii} & 3.10 &   --1.10   & --4.20  &  0.59   &	      & 	&	  &	   &	    & 0.08 &  1   &	    &	        &         &	  &	   &	       &	 &	  &      &   \\
\ion{Cr}{i}  & 5.64 &            &         &         &	      & 	&	  &	   &	    &	   &      &  1.48   &   <--4.16 &  <0.06  &	  &	   &	       &	 &	  &      & 1 \\
\ion{Ti}{ii} & 4.90 &    0.47    & --4.43  &  0.36   &  0.12  &	 0.61   & --4.29  &  0.11  &  0.11  & 0.09 &   2  &  1.24   &   --3.66  &  0.56   &  0.08 &  1.35  &   --3.55  &   0.38  & 0.11   & 0.13 & 5 \\
\ion{Fe}{i}  & 7.52 & 	 2.73    & --4.79  & 	     &  0.14  &	 3.12   & --4.40  &        &  0.13  & 0.14 &  27  &  3.30   &   --4.22  &         &  0.12 &  3.59  &   --3.93  &	 & 0.12   & 0.17 & 33 \\
\ion{Ni}{i}  & 6.23 &	 1.54    & --4.69  &  0.10   &        &	        &         &        &	    & 0.15 &   1  &  2.04   &   --4.19  &  0.03   &	  &	   &	       &	 &	  & 0.19 & 1 \\  
\ion{Sr}{ii} & 2.92 &   --2.38   & --5.30  & --0.51  &        &   --2.2 & --5.12  & --0.72 &	    & 0.09 &   1  & --1.24  &   --4.16  &  0.06   &  0.03 & --1.05 &   --3.97  &  --0.04 & 0.06   & 0.12 & 2 \\
\ion{Ba}{ii} & 2.17 &   <--2.80  & <--4.97 & <--0.18 &	      &	        &	  &        & 	    &      &   1  & <--2.1  & <--4.27   & <--0.05 &	  &	   &	       &	 &	  &      & 1 \\

\hline                                  
\end{tabular}
\end{table*}

\subsubsection{Medium-resolution analysis of ~\pritwo}

An estimate of the carbon content and metallicity of~\pritwo~is obtained for the medium-resolution GTC/OSIRIS spectrum by comparing theoretical spectra with the observed one. Model atmospheres were computed with the {\sc Kurucz} codes as described in \citet{meszaros12}. Modelled spectra were produced using synthesis code {\sc ASS$\epsilon$T}  \citep{asset}. We used the {\sc ferre}\footnote{Available from \url{http://github.com/callendeprieto/ferre}.} code \citep{ferre} to search for the best fit to the observed spectrum by simultaneously deriving the metallicity [Fe/H] and the carbon abundance [C/Fe] while keeping atmospheric parameters fixed to the values derived in ~\S~\ref{parameters}.

\subsection{NLTE Calculations}

At very low metallicities, non-local thermodynamic equilibrium (NLTE) effects are often significant \citep{asplund05}. The electron density decreases and the collision rates are greatly reduced with decreasing metal abundance. Furthermore, the radiation is absorbed by a smaller fraction of metal atoms and ions and the photoionization rate increases \citep{gehren04}. Since the
 departures of the excitation and ionisation states of the elements from the thermodynamic equilibria are more extreme in metal-poor and hotter atmospheres, the NLTE abundances provide a more realistic picture of the overall chemical pattern of our unevolved UMP targets.

As a first attempt to determine the NLTE abundance patterns that can be compared with Population III supernova nucleosynthesis yields, we performed the NLTE calculations for \ion{Na}{i}, \ion{Mg}{i}, \ion{Al}{i}, \ion{Si}{i}, \ion{Ca}{i-ii}, \ion{Ti}{i-ii}, \ion{Fe}{i-ii} , and \ion{Sr}{ii}, using the NLTE methods developed in our earlier studies and based on the most up-to-date atomic data available so far. In metal-poor atmospheres, collisional rates are mainly determined by collisions with neutral hydrogen atoms. For \ion{Na}{i} \citep{alexeeva_na}, \ion{Mg}{i} \citep{mash_mg13}, \ion{Al}{i} \citep{Baumueller_al1,mash_al2016}, \ion{Si}{i} \citep{2020MNRAS.493.6095M}, \ion{Ca}{i-ii}  \citep{mash_ca,2017A&A...605A..53M}, and \ion{Sr}{ii} \citep{1997ARep...41..530B,mash2021}, collisions with \ion{H}{i} atoms were treated by implementing the quantum-mechanical rate coefficients from calculations of \citet[][\ion{Na}{i}]{barklem2010_na}, \citet[][\ion{Mg}{i}]{mg_hyd2012}, \citet[][\ion{Al}{i}]{Belyaev2013_Al}, \citet[][\ion{Si}{i}]{Belyaev2014_Si}, \citet[][\ion{Ca}{i}]{2017ApJ...851...59B}, and \citet[][\ion{Sr}{ii}]{mash2021}. For \ion{Ti}{i-ii} and \ion{Fe}{i-ii}, we used the Drawinian rates \citep{Steenbock1984} scaled by $S_{\rm H}$=1 and 0.5, respectively, according to empirical estimates of \citet{sitnova_ti} and \citet{mash_fe}.
The coupled radiative transfer and statistical equilibrium equations were solved with the code {\sc detail} \citep{detail}, where the opacity package was revised as presented by \citet{mash_fe}.

For \ion{Li}{i}, we applied the NLTE abundance corrections from \citet{lind09}. In the literature, there are no NLTE corrections for lines of \ion{Sc}{ii}, \ion{Cr}{i}, and \ion{Ni}{i} for atmospheric parameters relevant to the analysed stars. Thus, we provide only LTE abundances for those elements.

Line-to-line elemental abundances corrected for NLTE are in Table~\ref{table:PR221EW}. The NLTE effects turn out to be rather small ($\Delta_{\rm NLTE}$ = \abu$_{\rm (NLTE)}$ -- \abu$_{\rm (LTE)}$ $\leq$ 0.05~dex) for \ion{Li}{i}, \ion{Na}{i}, and \ion{Ca}{ii}. On the other hand, significant abundance shifts were found for the minority species which are subject to overionization, such as \ion{Mg}{i}, \ion{Al}{i}, \ion{Ca}{i}, and \ion{Fe}{i}, with positive $\Delta_{\rm NLTE}$ corrections always larger than 0.3~dex.

\subsection{Comments on individual elements}

Lithium was detected in both stars, although we note that the continuum determination may have a significant impact on the measured abundances \citep[see also][]{starkenburg18}. 

We measured the 5890~\AA~line of \ion{Na}{i} for~\prione, being the same line in \pritwo~distorted by interstellar absorption. The blue Mg triplet and green \ion{Mg}{i} b lines were detected in both spectra and their resulting abundances are in good agreement. Al was measured from the \ion{Al}{i} 3961~\AA~line, since the other \ion{Al}{i} line at 3944~\AA~line is severely contaminated by CH features. 
It was possible to detect usable \ion{Si}{i} and \ion{Sc}{ii} lines only in the \prione~spectrum. 

For both objects, Ca abundances were derived from the \ion{Ca}{i} line at 4226~\AA. For~\prione~we also measured the calcium abundance from the synthesis of the \ion{Ca}{ii} K line. The abundances in LTE differ by 0.45 dex, with a higher \ion{Ca}{ii} abundance. However, the observed discrepancy is greatly reduced when abundances are corrected for NLTE. In \S~\ref{grav_check} we use ionisation equilibrium of calcium as a consistency check on the surface gravity value adopted for \prione~from photometry.

Abundances for \ion{Ti}{ii} were measured from two and five lines for \prione~and~\pritwo, respectively.
A \ion{Cr}{i} feature at 4254~\AA~can be detected in the spectrum of \pritwo, however such line is extremely weak and therefore extremely noisy and sensitive to the continuum placement. Thus, the provided value should be regarded as an upper limit to the actual \ion{Cr}{i} abundance measurement.
Abundances for \ion{Ni}{i} have been derived from the blue feature at 3858~\AA.
\ion{Sr}{ii} measurements are based on just one line in \prione~and two lines in \pritwo. 

In Table~\ref{table:abundances} we also provide upper limits for \ion{Y}{ii}, and \ion{Eu}{ii}. For these elements there is considerable uncertainty in the measurement of the EWs, because they are weak  (i.e. $<$ 10~m\AA) and noisy.

\subsection{Carbon and Nitrogen}

We use spectrum synthesis to analyse features of CH and NH and constrain the abundances of carbon and nitrogen. The synthetic spectra were computed with the {\sc synthe} code developed by Kurucz \citep{synthe} in its Linux version \citep{sbordone05} and adopting the same model atmospheres used to derive the abundances from EWs. The molecular line list are taken from \citet{masseron14} and the latest Kurucz compilation\footnote{\url{http://wwwuser.oat.ts.astro.it/castelli/linelists.html}}.

The C abundance is determined from the CH lines in the G band (the region between 4309-4314~\AA~and 4365-4369~\AA)~which are almost free from atomic blends. For ~\prione, further constraints on carbon abundances are obtained from the CH 3944 \AA\ line, which blends the \ion{Al}{i} resonance line at 3944~\AA. Such line was shown by \citet{arpigny83} to be severely blended with several CH lines, even in stars where CH was not a prominent feature of the spectrum. Corrections for the effect of evolutionary status on carbon are negligible (i.e. less than 0.01) for both stars, because of their unevolved nature \citep{placco14}.

An upper limit to the nitrogen abundance is derived for both stars from the ultraviolet NH band A$^{3} \Pi_{i}$ -- X$^{3}\Sigma^{-}$ at 3360~\AA.

\subsection{Consistency check on atmospheric parameters}\label{grav_check}

Calcium is observed in two ionisation stages in the spectrum of~\prione. From the \ion{Ca}{i} line at 4226~\AA, we measure an abundance of \abu~(Ca) = 1.77 $\pm$ 0.12, whereas the analysis of the  \ion{Ca}{ii} line at 3933~\AA ~gives \abu~(Ca) = 2.22 $\pm$ 0.14. Both lines are subject to departures from LTE, however, the NLTE corrections for the two lines are different. For the \ion{Ca}{i} line at 4226~\AA, NLTE raises the abundance by 0.3 dex and leads to an NLTE abundance of \abu~(Ca) = 2.07. Conversely, the NLTE correction for \ion{Ca}{ii} line at 3993~\AA~is slightly negative (--0.04 dex). This brings the abundances derived from lines of the two ionisation stages in good agreement. Thus, we conclude that the \ion{Ca}{i}/\ion{Ca}{ii} ionisation equilibrium is fulfilled for the atmospheric parameters we have adopted for \prione.

\begin{table}
\caption{Changes of the derived elemental abundances depending on stellar parameters.}             
\label{table:errors}    
\centering   
\footnotesize
\npdecimalsign{.}
\nprounddigits{2}
\nplpadding{1}

\begin{tabular}{r n{1}{2} n{1}{2} n{1}{2} n{1}{2}n{1}{2} n{1}{2}}       
\hline\hline  
\multicolumn{7}{c}{Pristine\_221.8781+09.7844}\\
\multicolumn{7}{c}{$\Delta$T$_{\rm eff}$= 134~K $\Delta \log$ g=0.1~dex $\Delta$v$_{\rm t}$= 0.5 km/s}\\
\hline
{Ion} &  {+$\Delta$ T$_{\rm eff}$} & {-$\Delta$ T$_{\rm eff}$}    & {+$\Delta\log$ g}  &  {-$\Delta\log$ g}  &    {+$\Delta$ v$_{\rm t}$}  &    {-$\Delta$ v$_{\rm t}$} \\
\hline
                 Fe~I	&    0.13 &   -0.131 &   -0.002  &   0.002 &	-0.034  &  0.044  \\
                 Li~I	&    0.10 &   -0.101 &   -0.003  &   0.002 &	 0.011  & -0.027  \\
                 Na~I	&    0.092 &   -0.096 &   -0.002  &   0.002 &	-0.008  &  0.007  \\
                 Mg~I	&    0.083 &   -0.085 &   -0.000  &   0.001 &	-0.023  &  0.026  \\ 
                 Al~I	&    0.103 &   -0.107 &   -0.001  &   0.001 &	-0.006  &  0.005  \\ 
                 Si~I	&    0.104 &   -0.106 &    0.001  &  -0.001 &	-0.009  &  0.009  \\ 
                 Ca~I	&    0.113 &   -0.116 &   -0.003  &   0.003 &	-0.039  &  0.051  \\ 
                 Ca~II  &    0.131 &   -0.138 &   -0.015  &   0.012 &   -0.016  &  0.007   \\  
                 Sc~II	&    0.073 &   -0.076 &    0.032  &  -0.032 &	-0.005  &  0.006  \\ 
                Ti~II	&    0.066 &   -0.068 &    0.032  &  -0.032 &	-0.035  &  0.049  \\ 
                Ni~I	&    0.134 &   -0.137 &   -0.001  &   0.001 &	-0.006  &  0.005  \\ 
                Sr~II	&    0.079 &   -0.081 &    0.032  &  -0.031 &	-0.006  &  0.009  \\ 
\hline
                 	&     &   &     &   &	  &    \\
                 	&    &    &     &    &	   &   \\

\hline
\hline
\multicolumn{7}{c}{Pristine\ 237.8588+12.5660}\\
\multicolumn{7}{c}{$\Delta$T$_{\rm eff}$= 175~K $\Delta \log$ g=0.1~dex $\Delta$v$_{\rm t}$= 0.5 km/s}\\
\hline
{Ion} &  {+$\Delta$ T$_{\rm eff}$} & {-$\Delta$ T$_{\rm eff}$}    & {+$\Delta\log$ g}  &  {-$\Delta\log$ g}  &    {+$\Delta$ v$_{\rm t}$}  &    {-$\Delta$ v$_{\rm t}$} \\
\hline

                Fe~I	&    0.168 &   -0.171 &   -0.002  &   0.002 &	-0.026  &  0.045  \\  
                Li~I	&    0.130 &   -0.134 &   -0.002  &   0.003 &	 0.014  & -0.029  \\  
                Mg~I	&    0.113 &   -0.118 &   -0.001  &   0.002 &	-0.026  &  0.027  \\  
                Al~I	&    0.139 &   -0.144 &    0.000  &   0.001 &	-0.004  &  0.003  \\  
                Ca~I	&    0.152 &   -0.157 &   -0.002  &   0.002 &	-0.028  &  0.033  \\  
                Ti~II	&    0.085 &   -0.089 &    0.033  &  -0.033 &	-0.062  &  0.093  \\  
                Cr~I	&    0.173 &   -0.178 &   -0.002  &   0.002 &	-0.005  &  0.005  \\  
                Ni~I	&    0.179 &   -0.186 &   -0.001  &   0.001 &	-0.006  &  0.005  \\  
                Sr~II	&    0.109 &   -0.111 &    0.031  &  -0.031 &	-0.020  &  0.029  \\  

\hline
                                
\end{tabular}
\end{table}

\section{Results}\label{results}

\subsection{Constraints on the C and N abundances}\label{const_on_carbon}

\citet{starkenburg18} were able to put an upper limit of \abu(C) $\leq$ 5.6 to the carbon abundance of \prione~using the WHT/ISIS spectrum at intermediate resolution (R$\sim$2400). Despite the relatively good SNR of their high-resolution UVES spectrum (SNR$\sim$~45~at~4000~\AA), the quality of the data was not sufficient to detect carbon enhancement below the level already set by the higher SNR medium-resolution spectrum. 

\begin{figure*}
\centering
\includegraphics[width=0.98\textwidth]{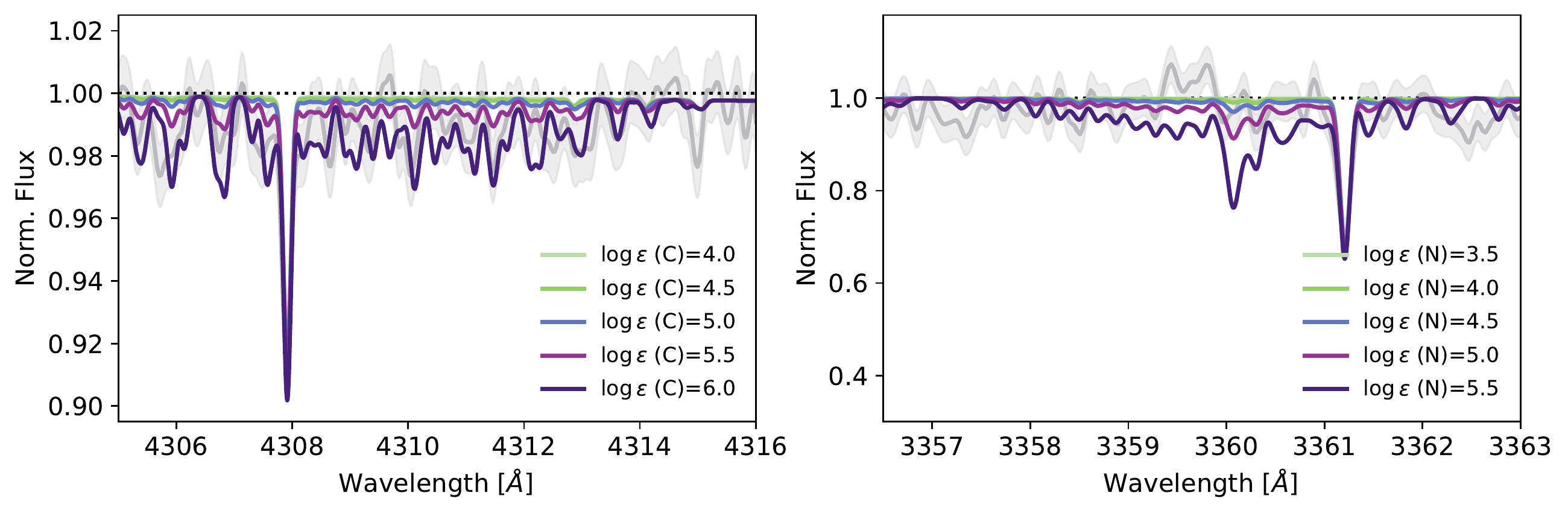}
      \caption{The comparison of the UVES high-resolution spectrum of~\prione~(thick grey line)
      with model spectra with with T$_{\rm eff}$ = 5683 K, $\log$(g) = 3.03, [M/H] = --4.79 and varying carbon and nitrogen abundances (see legend) is shown in the left- and right-hand panel, respectively. The noise level is represented by the grey shaded area}. All synthetic spectra are smoothed to a resolution of R = 30\ 000.
\label{fig:carbon}
\end{figure*}

\begin{figure*}
\centering
\includegraphics[width=0.7\textwidth]{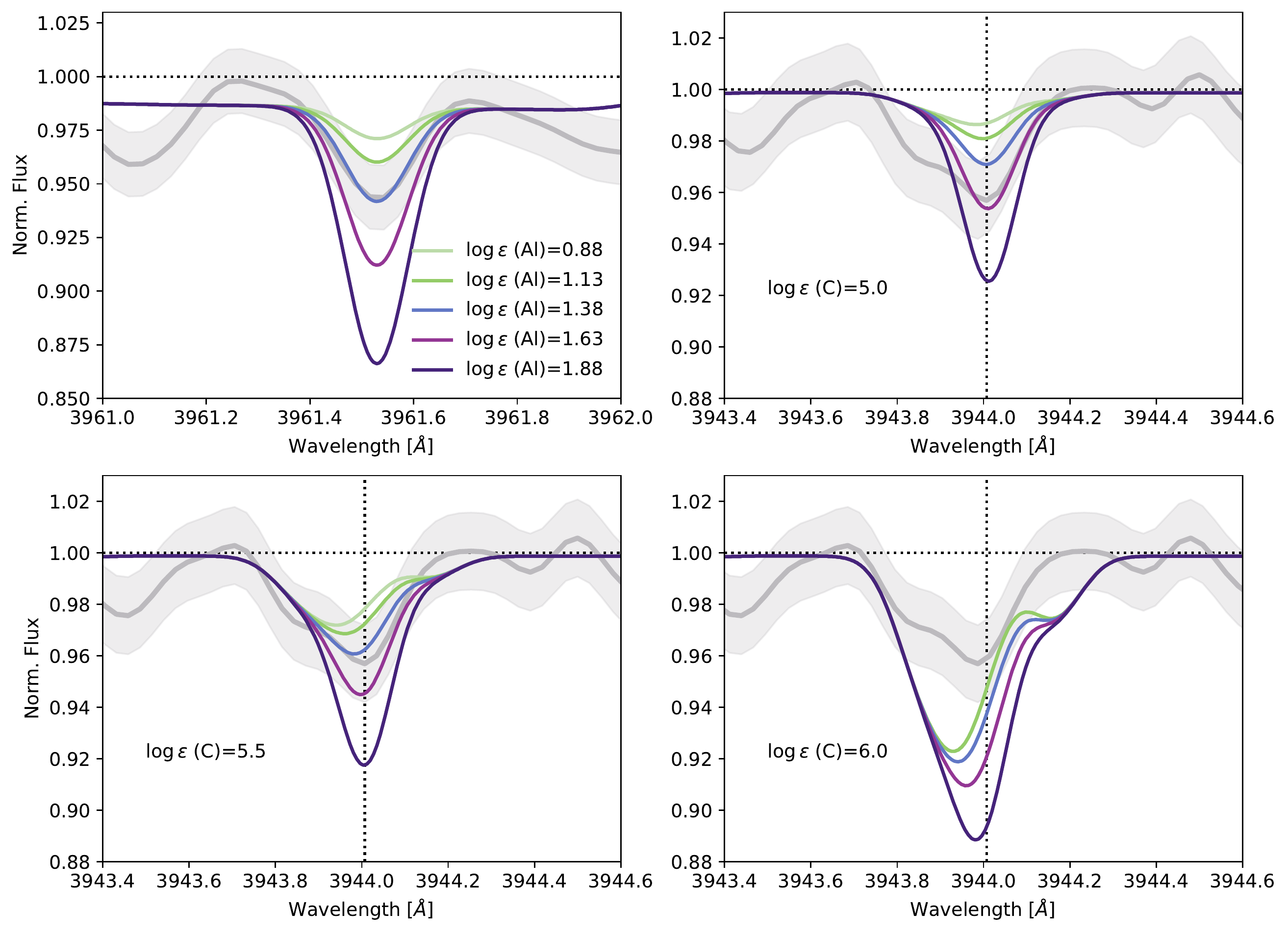}
      \caption{The top-left panel shows synthetic spectra with (from top to bottom) \abu (Al)= 0.88, 1.13, 1.38, 1.63, and 1.88 around the \ion{Al}{i} line at 3961~\AA. The effects of changing aluminium abundances by the same amounts for the modelled spectra around the \ion{Al}{i} 3944~\AA~line are plotted for different carbon abundances $\log \epsilon$ (C) = 5.0, 5.5, and 6.0 (top-right, bottom-left, and bottom-right, respectively).
      All modelled spectra are smoothed to a resolution of R = 30\ 000.
      The observed spectrum of~\prione~is shown as a grey thick line in all panels. The noise level is represented by the grey shaded area}.
\label{fig:aluminum}
\end{figure*}

Because of the low metallicity and the relatively high effective temperatures of the target stars, large changes in  carbon abundances have an extremely small impact on the appearance of the (weak) CH absorption. 
The synthetic spectra of Figure~\ref{fig:carbon} show that the difference between \abu(C) = 5.5 and \abu(C) = 6 never exceeds the 2\% of the flux. Despite the noise and the uncertainties in the continuum placement, it seems that the theoretical spectrum with \abu (C) $\leq$ 6.0 ([C/Fe] $\leq$ +2.29) presents a conservative upper limit to the carbon abundance. However, the SNR of the combined UVES spectrum of \prione~is still too low to permit accurate analysis of individual CH lines at the wavelengths shown in Fig.~\ref{fig:carbon}.



As noted by \citet{arpigny83}, the \ion{Al}{i} line at 3944~\AA~is severely contaminated by CH. We used the 3944~\AA~blend to check the estimated C abundance of~\prione. The aluminium NLTE and LTE abundances were fixed from analysis of the second line of the resonance doublet, \ion{Al}{i}~3961~\AA. Since the NLTE abundance corrections for \ion{Al}{i} 3944 and 3961~\AA~are equal, in this section, we perform analysis of these lines under the LTE assumption. Figure~\ref{fig:aluminum} (top left-hand panel) shows that \abu (\ion{Al}{i}) = 1.38 represents the best LTE fit to the observed spectrum of \ion{Al}{i} 3961~\AA.

The same figure~\ref{fig:aluminum} also shows the impact of  different carbon abundances on the synthetic spectra computed around the \ion{Al}{i} line at 3944~\AA. An increasing abundance of carbon mainly enhances the spectral feature at the blue end of the \ion{Al}{i} 3944~\AA~line. If one assumes that the blue wing is noise and that \abu (C)=5.0 (top right panel of Fig.~\ref{fig:aluminum}), we derive an abundance of \abu (\ion{Al}{i}) = 1.54 
, which is greater than the value measured from the fit of the \ion{Al}{i} line at 3961~\AA . This indicates, that the solution with \abu~(C) = 5.0 is not favoured, since carbon should be more abundant to contribute to the left wing of the \ion{Al}{i}~3944~\AA~blend. 

The model with $\log \epsilon$ (C)=5.5 and \abu (\ion{Al}{i}) = 1.38 seems to reproduce well the observed \ion{Al}{i}  3944~\AA~feature, and in this case the derived Al abundance agrees with that inferred from the clean~\ion{Al}{i}~3961~\AA. The \abu~(C)= 5.5 value is lower than the C abundance deduced from the G band and supports our conclusion that \prione~does not belong to strongly C-enhanced stars.


\begin{figure*}
\includegraphics[width=0.98\textwidth]{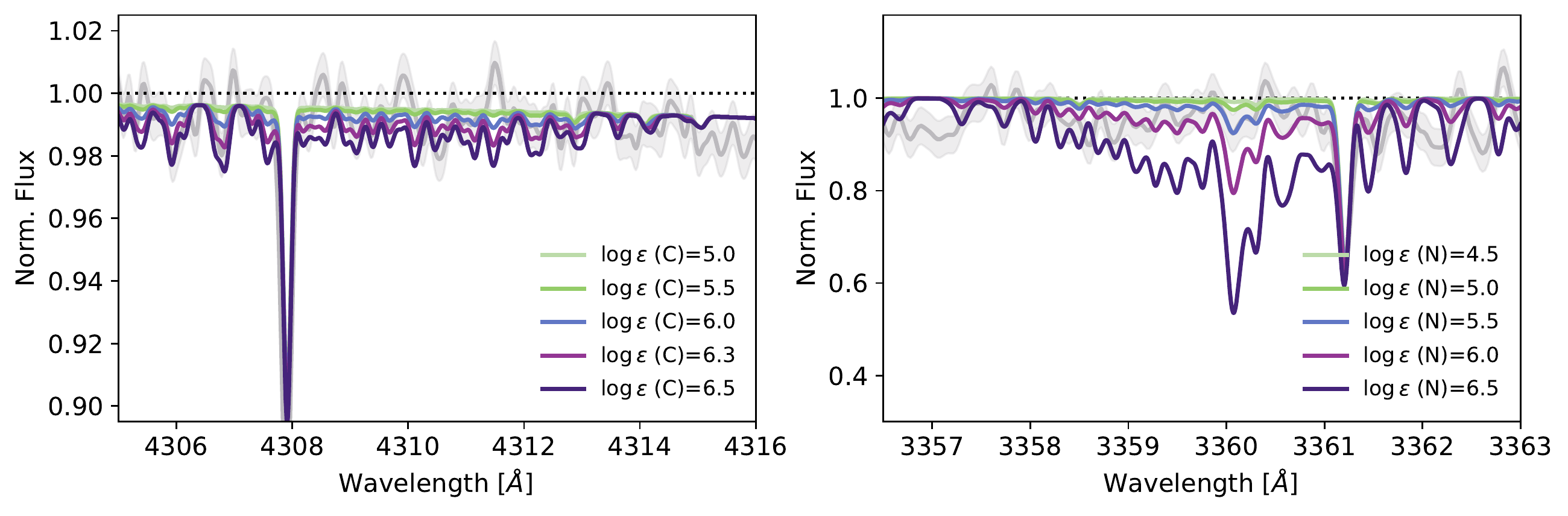}
      \caption{The same as for Figure~\ref{fig:carbon} for~\pritwo.}
\label{fig:pr237_CN}
\end{figure*}

Figure~\ref{fig:carbon} shows the observed spectrum of \prione~around the NH band at 3360~\AA. From the comparison of the observed spectrum with synthetic ones with varying nitrogen abundances, we estimate an upper limit of \abu~(N) $\leq$ 5.0 to the nitrogen content of~\prione.

In Figure~\ref{fig:pr237_CN}, we compare the UVES spectrum of~\pritwo~with theoretical models with varying carbon and nitrogen abundances. 
Because of the high effective temperature and low metallicity, models spanning a large range in carbon abundances (from $\log \epsilon$(C)=5.0 to 6.5) differ by less than 2\% of the flux. Thus, despite the high quality of the data, we are only able to put an upper limit to the C abundance of \abu~(C) < 6.3. This would correspond to to [C/Fe] < +2.02, when taken together with the high-resolution measurement of iron, [Fe/H]= --4.22. 
A similar upper limit of [C/Fe] < 1.90 can be derived by the medium resolution spectrum using metallicity from the high-resolution UVES spectrum. When the measured carbon abundance is taken together with the metallicity derived from {\sc ferre} from the medium-resolution data, the upper limit to the carbon abundance is [C/Fe]< +2.10 (see also Figure~\ref{fig:ferre}).Therefore, we assume \abu~(C) <  6.3 as a conservative upper limit to the carbon abundance of ~\pritwo.
From a careful inspection of the right-hand panel of Figure~\ref{fig:pr237_CN}, we adopt \abu~(N) $\leq$ 6.0 as the upper limit for the nitrogen abundance of ~\pritwo.

Due to the reduced radiative heating with a shortage of spectral lines, the temperature structures of metal-poor 3D hydrodynamical model atmospheres are much cooler in the higher atmospheric layers \citep[e.g.][]{asplund05}. This translate into large 3D effects for temperature-sensitive species like molecules \citep[see  e.g.][]{asplund2001,gonzalez10,gallagher16,gallagher17,nordlander17,collet07,collet18}.
The complete 3D analysis of target stars is beyond the scope of this study, but we refer to the 3D corrections derived by \citet{caffau12} for SDSS J102915+172927 -- a star with stellar parameters and abundances similar to ~\prione~\citep[see discussion in][]{starkenburg18} -- to give a qualitative idea of the  corrections expected 
for the carbon and nitrogen abundances discussed here. The work of \citet{caffau12} shows that the use of 3D, LTE modelling (as opposed to our 1D, LTE procedure) would lead to significantly lower abundances for both carbon and nitrogen (by --0.7 and --0.9 dex, respectively). Unfortunately, no detailed calculations for departures from LTE for CH and NH with realistic molecular models exist.

\subsection{Overall chemical pattern}\label{literature}
We measured the equivalent widths of 27 \ion{Fe}{i} lines for the combined spectrum of \prione, which yield a final LTE abundance of [\ion{Fe}{i}/H]$_{\rm LTE}$=--4.79 $\pm$ 0.14. This corresponds to [\ion{Fe}{i}/H]$_{\rm NLTE}$=--4.40 $\pm$ 0.13. From the analysis of 33 \ion{Fe}{i} lines in the UVES spectrum of ~\pritwo, we measure [\ion{Fe}{i}/H]$_{\rm LTE}$=--4.22 $\pm$ 0.12 and [\ion{Fe}{i}/H]$_{\rm NLTE}$=--3.93 $\pm$ 0.12. The medium-resolution analysis of the GTC/OSIRIS spectrum returns a higher metallicity of [Fe/H]=--3.86 $\pm$ 0.20 dex.

\begin{figure}
\centering
\includegraphics[width=\columnwidth]{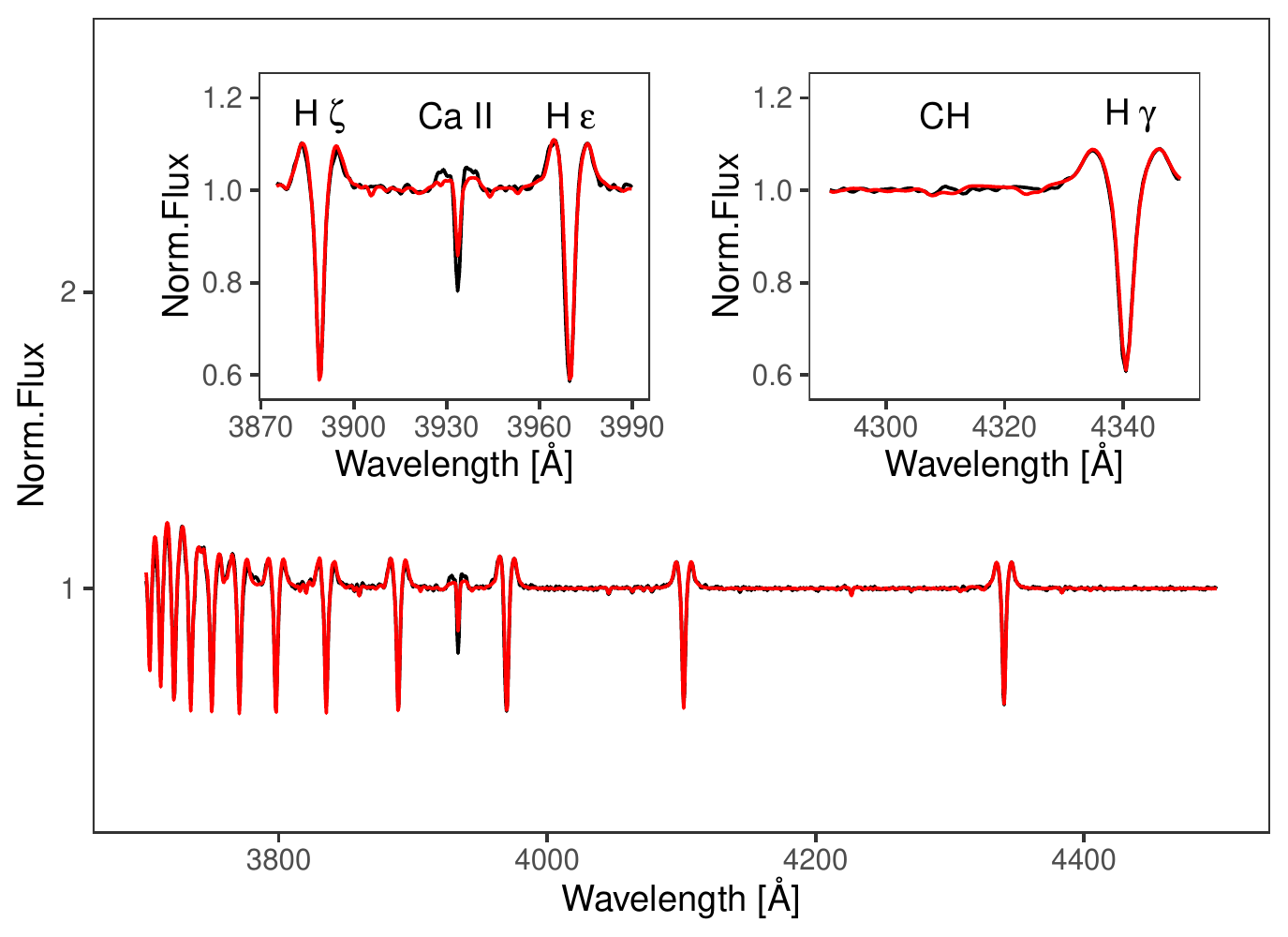}
      \caption{The medium-resolution spectrum for \pritwo~as observed with GTC/OSIRIS and normalised by a running mean by the {\sc ferre} code is shown in black. The best-fit spectrum --with the same normalisation-- is overplotted in red. The insets display a zoom around the \ion{Ca}{ii} K line and the G-band.}
\label{fig:ferre}
\end{figure}

Figure~\ref{fig:ferre} shows the observed GTC/OSIRIS spectrum normalised by a running mean by {\sc ferre} along with the best-fit model --with the same normalisation \citep{agauado17a,aguado17b}. The metallicity [Fe/H]=--3.86 dex derived from the GTC/OSIRIS spectrum is in good agreement with what is found from the high-resolution UVES analysis under the NLTE assumption ([Fe/H]$_{\rm NLTE}$=--3.93+-0.12). This is because -- due to the extreme weakness of most of the absorption lines-- most information for the metallicity determination in the intermediate resolution data comes from the \ion{Ca}{ii} K line, for which NLTE corrections are extremely small ($\sim$ --0.02 dex). However, we also note that the \ion{Ca}{ii} K is blended with with an interstellar Ca absorption line in the GTC/OSIRIS spectrum. This may also account for the higher metallicity derived from intermediate-resolution spectrum relative to the UVES one.

In the low metallicity regime, and increasing fraction of MW halo stars has been found to be greatly enhanced in carbon, the so-called carbon-enhanced metal-poor (CEMP) stars \citep[e.g.,][]{beers05}.
Depending on their abundance pattern, CEMP stars have been traditionally classified in several sub-classes: 
CEMP-s, CEMP-r, CEMP-r/s, and CEMP-no \citep{beers05}\footnote{CEMP-s : [C/Fe] $>$ +1.0, [Ba/Fe] $>$ +1.0, and [Ba/Eu] $>$ +0.5; CEMP-r: [C/Fe] $>$ +1.0 and [Eu/Fe] $>$ +1.0; CEMP-r/s: [C/Fe] $>$ +1.0 and 0.0 $<$ [Ba/Eu] $<$ +0.5 CEMP-no: [C/Fe] $>$ +1.0 and [Ba/Fe] $<$ 0.0.}.
The high carbon enhancement accompanied by strong over-abundances of neutron-capture elements produced by the main $s$-process observed in both CEMP-s and CEMP-r/s stars is thought to be produced by mass-transfer from low- to intermediate-mass (from 1 to 4 M$_{\odot}$) asymptotic giant-branch (AGB) companion \citep[e.g.][]{starkenburg14,hansen15,abate15,sakari18,choplin21}.

On the contrary, the carbon enhancement with low (or absent) abundances of neutron-capture elements observed in CEMP-no stars is thought to reflect directly the composition of the natal gas cloud from which the star first formed. 
Several astrophysical sites have been proposed for the progenitors of these bona-fide second generation stars; including 
``faint supernovae (SNe)'' that undergo mixing and fallback  \citep[e.g.,][]{tominaga14,komiya20}, rapidly rotating massive stars of ultra low metallicity \citep[e.g.][]{meynet06,chiappini08,maeder15,choplin16}, and 
metal-free massive stars \citep{heger10}.

\begin{figure}
\centering
\includegraphics[width=0.7\columnwidth]{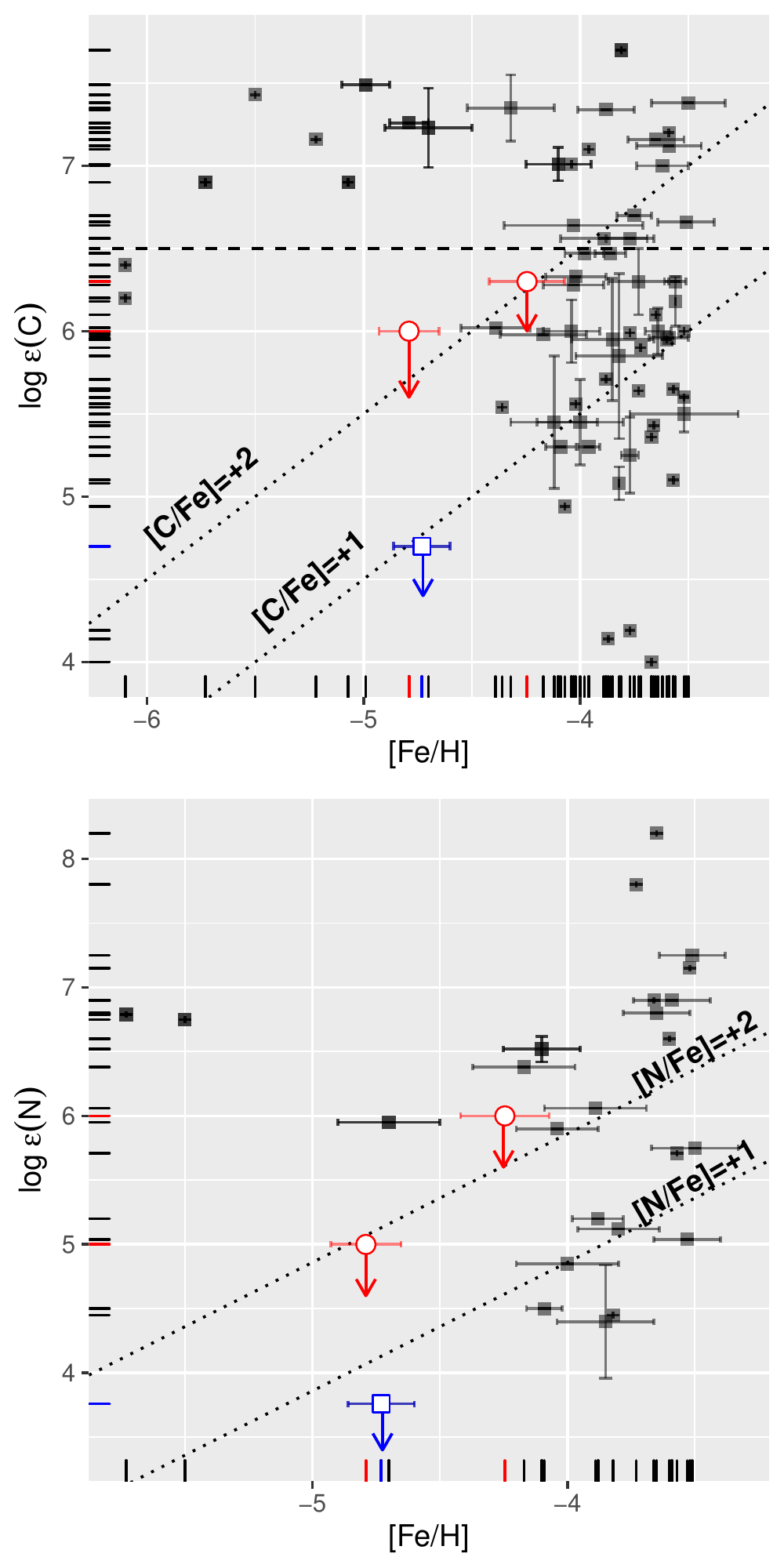}
      \caption{C and N abundances for the two stars compared to literature values. Red empty circles are the two targets analysed in this study, whereas the blue empty square represents SDSS\ J102915+172927 \citep{caffau11}, the most metal-poor star observed so far with a similar overall chemical pattern to ~\prione~\citep[see][for a discussion]{starkenburg18}. Stars from literature with [Fe/H] $\leq$ --3.5 and $\log$ (g) $\geq$ 3 are shown as black symbols. The horizontal dashed line in the top panel gives the mean \abu~(C) = 6.5 value  used by \citet{spite13} to define carbon enhanced metal-poor stars at metallicities [Fe/H] < --3.4.}
\label{fig:carbon_nitrogen}
\end{figure}

In Figure~\ref{fig:carbon_nitrogen} [C/Fe] and [N/Fe] LTE abundances are shown as a function of [Fe/H] for~\prione~and~\pritwo~and  unevolved stars (e.g., $\log$g$\geq$3) with metallicities lower than [Fe/H]<--3 collected from the SAGA database\footnote{Available at \url{http://sagadatabase.jp}}. In the figure, \prione~and~\pritwo~are represented as empty red circles, whereas data from the literature are shown as black filled squares. In the same figure we also highlight SDSS\ J102915+172927 (shown as empty blue square; \citealp{caffau11}),
the star  with lowest metallicity known.
Inspection of the top panel of Figure~\ref{fig:carbon_nitrogen} leads us to conclude that both~\prione~and~
\pritwo~belong to the group of low-carbon band \citep{bonifacio15} with both stars possibly being ``carbon-normal''\footnote{\citet{bonifacio15} define ``carbon-rormal'' stars with [Fe/H]<--4 and \abu(C)$\leq$ 5.5 or [Fe/H]$\geq$--4 and [C/Fe] < +1.0. Low-carbon band are stars that do not fulfil the ``carbon normal'' criterion and have \abu(C) $\leq$ 7.6.}.
Both stars are classified as CEMP-no stars, as they do not show any overabundance of neutron-capture elements (see ~\S~\ref{literature}) and they are associated with the low-C region (\abu~(C) $\leq$ 7.1; \citealp{yoon16}). 
Additionally, among the CEMP-no sample itself, two very different behaviours in the~\abu(C) versus [Fe/H] space can be observed and they are likely associated to different classes of progenitors \citep{yoon16}. In particular, both stars  would be classified as Group II CEMP-no stars (e.g., stars for which a clear dependence of \abu~(C) on [Fe/H] is observed, see top panel of Figure~\ref{fig:carbon_nitrogen})\footnote{This conclusion is further supported by the measured abundance of the light-elements Na and Mg in the two stars (see Figure 4 in \citet{yoon16}).} and they can be tentatively associated faint mixing-and-fallback SNe progenitors \citep{nomoto13}.

Similarly to carbon, also nitrogen appears to be underabundant in~\prione~when compared to stars of similar metallicity. However, only a few measurements of N abundance exist at this low-metallicity regime to draw firm conclusions.

\begin{figure*}
\includegraphics[width=0.95\textwidth]{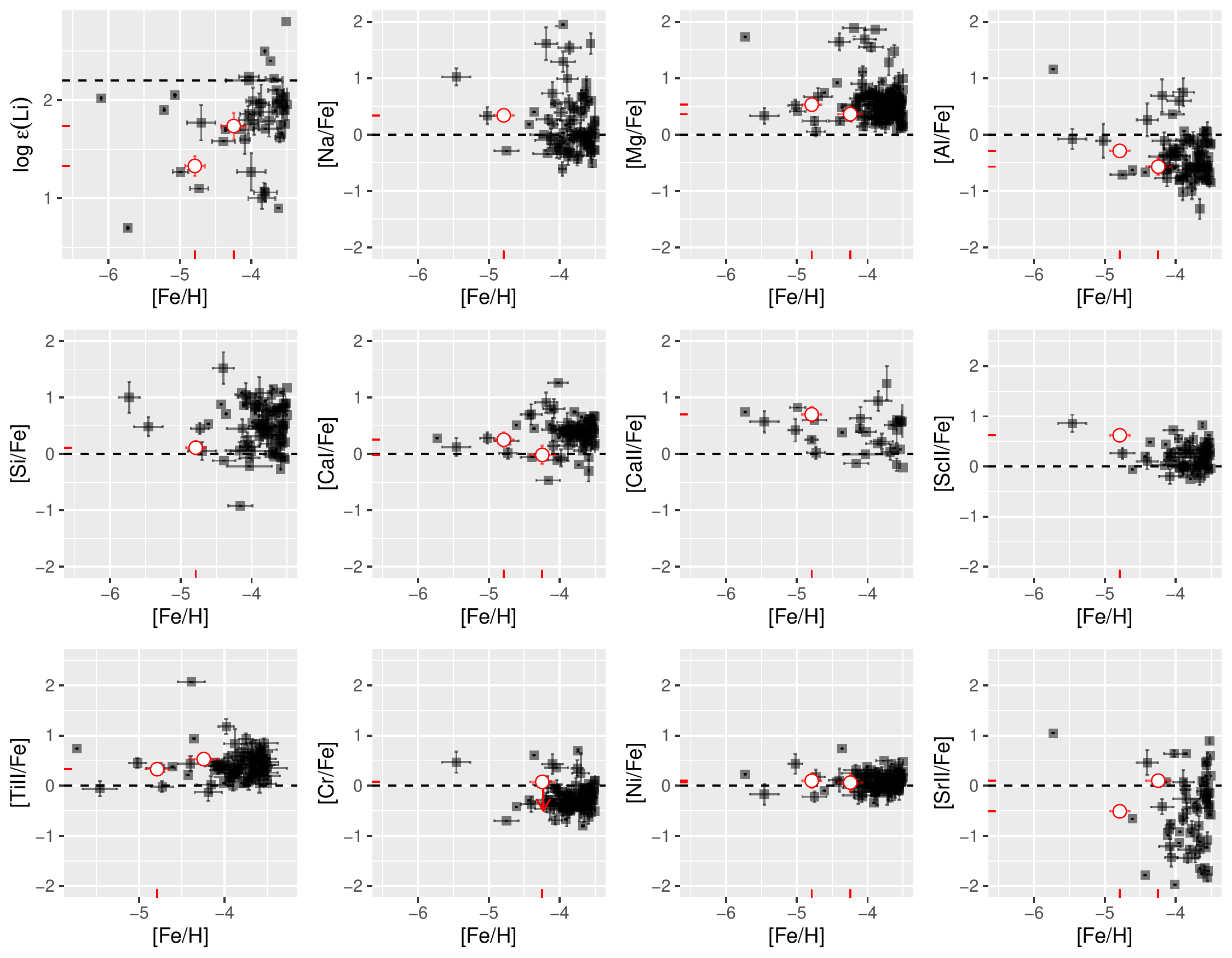}
      \caption{LTE Li abundances (\abu~(Li) and [X/Fe] abundance ratios for 11 representative elements in \prione~and~\pritwo~(shown as red emty symbols) are compared to literature values (black squares).}
\label{fig:elements}
\end{figure*}

In Figure~\ref{fig:elements} the absolute abundance of Li (\abu~(Li)), as well as abundance ratios [X/Fe] for 11 representative elements (X) in LTE  are shown as a function of [Fe/H], for~\prione,~\pritwo,  and stars with metallicities lower than [Fe/H]<--3 taken from the SAGA database\footnote{Aivalable at \url{http://sagadatabase.jp}}. In this figure, \prione~and~\pritwo~are represented as empty red circles, whereas data from literature are shown as black filled squares. 


\begin{figure}
\centering
\includegraphics[width=0.88\columnwidth]{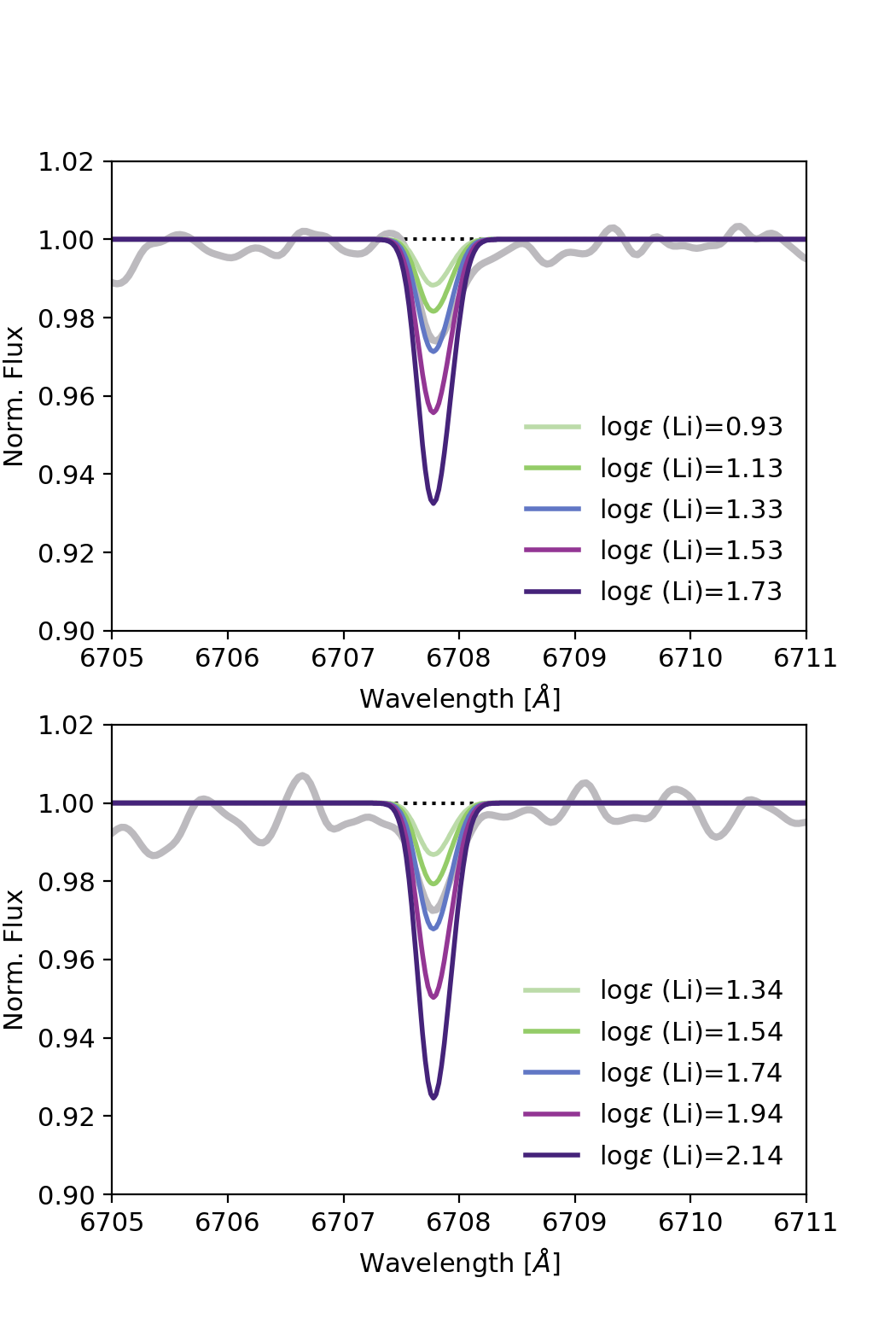}
      \caption{The \ion{Li}{i} resonance doublet in \prione~(upper panel) and \pritwo~(lower panel) are shown in grey. Theoretical spectra with varying lithium abundance (see legend) are also shown.}
\label{fig:lithium}
\end{figure}

The measured lithium abundances is \abu~(\ion{Li}{i})$_{\rm LTE}$ = 1.33 and \abu~(\ion{Li}{i})$_{\rm LTE}$ = 1.74 for~\prione~and~\pritwo, respectively. The uncertainties we derive from the EW measurements are extremely small for both objects (0.02 dex), but a more realistic determination of the uncertainties associated to \ion{Li}{i} measurements of 0.2 dex can be obtained taking into account uncertainties due to the continuum placement \citep[e.g.][]{starkenburg18,kielty20}.
Figure~\ref{fig:lithium} shows synthetic spectra around the \ion{Li}{i} resonance line for ~\prione~and~\pritwo. The best fit solution is shown along with synthetic spectra with \ion{Li}{i} abundance changed by $\Delta$ \abu~(\ion{Li}{i}) = $\pm$ 0.2 and 0.4. 
The top left-hand panel of Figure~\ref{fig:elements} we compare the absolute Li abundances derived for the two stars with \abu~(Li) values form the literature for stars with surface gravities $\log$ g > 3. 
Both stars have \ion{Li}{i} abundance below the plateau found by \citet{spite82a,spite82b} for old and unevolved stars. This is in line with observations of stars with [Fe/H] $\leq$ --3  \citep{sbordone10}, with the notable exception of the dwarf star J0023+0307 with an [Fe/H] < --6.1 and a Li abundance of \abu(\ion{Li}{i}) = 2.02$\pm$0.08 dex \citep{aguadoLI}. For ~\prione~the observed \ion{Li}{i} abundance can also be ascribed to its relatively low effective temperature.
We refer to \citet{gonzalez19,gonzalez20} for a complete discussion on Li abundances in EMP/UMP stars.

The [Na/Fe] and [Al/Fe] LTE ratios appear to vary from star to star by a very large factor at lower metallicities (see Figure~\ref{fig:elements}). The scatter of data for stars of close metallicity is, in part, due to neglecting the departures from LTE. For [Fe/H] $< -3$, the Na and Al abundances can only be derived
from resonance lines, for which the NLTE effects are strongly dependent on atmospheric parameters and  element abundance itself 
\citep{1998A&A...338..637B,1997A&A...325.1088B}.
For~\prione, the NLTE abundance of Na is found to be lower than the LTE abundance, by 0.40~dex, while, in contrast, the Al abundance is higher in NLTE than in LTE, by 0.68~dex. NLTE correction of +0.60~dex was calculated for \ion{Al}{i} 3961~\AA\ in \pritwo. With the NLTE effects taken into account, we retrieve [Na/Fe] and [Al/Fe] abundance ratios close to their solar values, in line with available NLTE studies of the EMP stars \citep{andrievsky07,bonifacio09,2017A&A...608A..89M}  and the predictions by \citet{samland98} for the Galactic [Na/Fe] versus [Fe/H] trend.



Our UMP stars~\prione~and~\pritwo~appear to be typical Galactic halo stars with respect to the $\alpha$-elements Mg, Ca, and Ti, and also Si in Pr~221 (see Figure~\ref{fig:elements}). Magnesium is enhanced relatively to Fe, with [Mg/Fe]$_{\rm LTE}$ = 0.53 and 0.26 in \prione~and~\pritwo, respectively. Slightly lower abundance ratios are derived in NLTE, with [Mg/Fe]$_{\rm NLTE}$ = 0.46 and 0.24, respectively. In~\prione, calcium is measured in two ionisation stages. The abundance discrepancy found in LTE between \ion{Ca}{i} and \ion{Ca}{ii} nearly disappears in NLTE (see ~\S~\ref{grav_check}), and we derive a moderate enhancement of Ca ([\ion{Ca}{i}/Fe]$_{\rm NLTE}$ = 0.14 and [\ion{Ca}{ii}/Fe]$_{\rm NLTE}$ = 0.24). For~\pritwo, the [Ca/Fe] ratio is around
solar, with [\ion{Ca}{i}/Fe]$_{\rm LTE}$ = --0.06 and [\ion{Ca}{i}/Fe]$_{\rm NLTE}$ = --0.09. Titanium exhibits a substantial overabundance of Ti relatively to Fe in the LTE analysis, with [Ti/Fe]$_{\rm LTE}$ = 0.36 and 0.56 for~\prione~and~\pritwo, respectively. Since the NLTE corrections for lines of \ion{Ti}{ii} are smaller than that for lines of \ion{Fe}{i}, we obtain lower abundance ratios of [\ion{Ti}{ii}/Fe]$_{\rm NLTE}$ = 0.11 and 0.38 for~\prione~and~\pritwo, respectively. Thus, in~\prione, $\alpha$-enhancement is stronger for Mg than for Si, Ca, and Ti. 



Our UMP stars~\prione~and~\pritwo~are not different from the Galactic EMP stars also with respect to Sc, Ni, and probably Cr (see Figure~\ref{fig:elements}).
The scandium abundance was measured for~\prione~only, from one \ion{Sc}{ii} line at 4246~\AA. We derived a scandium enhancement at the level of [Sc/Fe]$_{\rm LTE}$ = +0.59. Nickel follows iron. The [Ni/Fe]  abundance ratio is expected to be nearly free of the NLTE effects because abundances of both elements were derived from lines of their neutral species, which have probably similar departures from LTE  \citep{2017A&A...608A..89M}. For Cr, we could measure only the upper limit and in one star, Pr~237.



We measure different [Sr/Fe] abundance ratios forfor~\prione~and~\pritwo, with [Sr/Fe]$_{\rm NLTE}$= -0.72 ([Sr/Fe]$_{\rm LTE}$ = --0.51) and [Sr/Fe]$_{\rm NLTE}$ = --0.04 ([Sr/Fe]$_{\rm LTE}$ = +0.06), respectively. Again, this is not exceptional for the extremely metal-poor regime of our Galaxy (see Figure~\ref{fig:elements}). It was not possible to detect spectral features of any other neutron-capture element. Taken
together with the measured low [Sr/Fe] ratios, this evidence would indicate that in both stars heavy elements are likely deficient with respect to iron. To clarify this issue, we compared theoretical spectra computed for varying Ba abundance with the observed spectra around the \ion{Ba}{ii} 4934~\AA\ line.
From this comparison, we conclude that if Ba would overabundant relative to Fe at the level [Ba/Fe] = +0.5, we would have observed a clear \ion{Ba}{ii} absorption line in the UVES spectra of the two targets. Thus,
we estimated an upper limit (in LTE) of~\abu~(Ba) $\leq$ --2.8 and  and $\leq$ --2.1 for ~\prione~and~\pritwo, respectively (Table~\ref{table:abundances}).


\begin{figure}
\includegraphics[width=0.95\columnwidth]{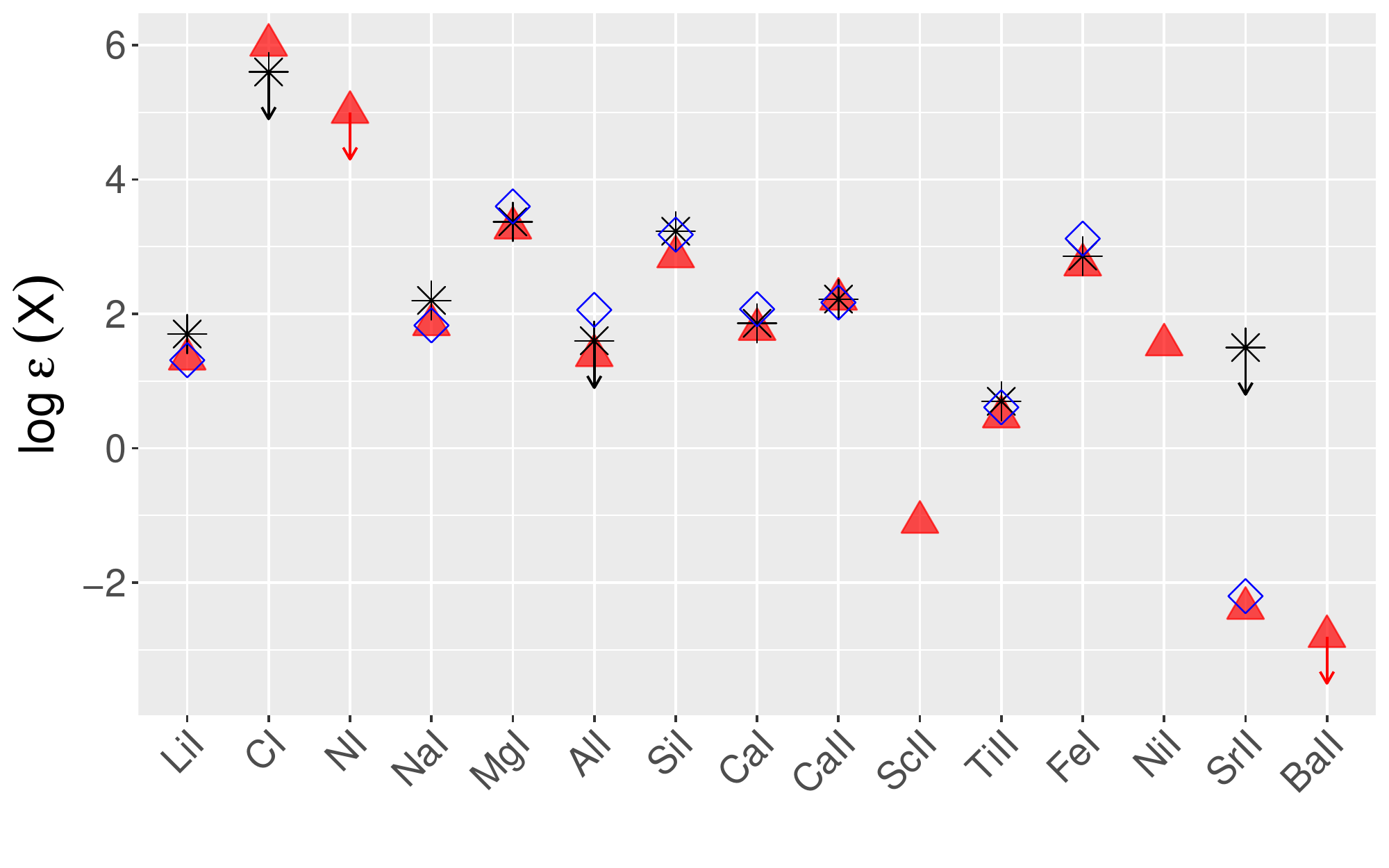}
      \caption{LTE (filled red triangles) and NLTE (empty blue diamonds) abundances derived in this study for ~\prione~are compared with the LTE abundances from \citet{starkenburg18} (shown as black asterisks).}
\label{fig:comparison}
\end{figure}

\subsection{Comparison with previous spectroscopic studies}

The abundance of iron determined from \ion{Fe}{i} lines in this study agrees with the value published \citet{starkenburg18} (\abu~(\ion{Fe}{i}) = 2.86 $\pm$ 0.13) within the measurement errors. \citet{starkenburg18} have adopted a T$_{\rm eff}$=5792~K, which is $\sim$100~K warmer than the one used here. According to Table~\ref{table:errors}, the iron abundance from \ion{Fe}{i} would increase by $\sim$0.13 when T$_{\rm eff}$ is raised by 134~K. This is the exact \ion{Fe}{i} difference measured in the two analysis. Therefore, we conclude that the difference in the derived iron abundances can entirely be attributed to the stellar parameters adopted in the two studies. 
Figure~\ref{fig:comparison} shows that our LTE measurements are generally in line with those presented in \citet{starkenburg18} for the elements in common in the two studies (e.g., they are consistent within the uncertainties associated to the measurements). 

The higher signal-to-noise ratio of the combined UVES spectrum of~\prione~allows us  to also provide measurements for elements for which only upper limits were derived in \citet{starkenburg18}. From the \ion{Al}{i} line at 3961~\AA~we derive a LTE abundance of \abu~(\ion{Al}{i})$_{\rm LTE}$ = 1.38, which corresponds to \abu~(\ion{Al}{i})$_{\rm NLTE}$= 2.06. This value is in good agreement with the upper limit of \abu~(\ion{Al}{i})$_{\rm LTE} \leq$1.6 measured by \citet{starkenburg18}. Finally, \citet{starkenburg18} provide the upper limit of \abu~(\ion{Sr}{ii}$_{\rm LTE}$ $\leq$--1.5 for the fit of the \ion{Sr}{ii} line at 4077.7~\AA. From the same line, we measure an abundance of \abu~(\ion{Sr}{ii})$_{\rm LTE}$  = --2.38 in LTE, which correspond to \abu~(\ion{Sr}{ii})$_{\rm NLTE}$  = --2.20 in NLTE. Finally, we present abundances (and upper limits) for nitrogen, scandium, nickel, and barium  which were not measured in \citet{starkenburg18}.

 For~\pritwo~and the elements in common with \citet{kielty20} we observe a perfect agreement between their abundances and our measurements. From the analysis of five \ion{Fe}{i} and three \ion{Fe}{ii} lines, they were able to put an upper limit of [Fe/H] $\leq$ --4.26, which is consistent with the value we have measured from the analysis of 33 lines ([\ion{Fe}{i}/H]$_{\rm LTE}$=--4.22 $\pm$ 0.12). For Li, they measure \abu~(\ion{Li}{i})=1.54 $\pm$ 0.20 which is still compatible with the abundance provided by our analysis given the large uncertainty introduced by the continuum placement when fitting the weak Li feature.  \citet{kielty20} measured \abu~(\ion{Mg}{i}) = 3.43 $\pm$ 0.13, which again agrees with the magnesium abundance of Table~\ref{table:abundances}.

\subsection{Model Predictions for UMP progenitor supernovae}\label{sec:progenitors}

We tentatively assess the properties of the progenitor population of \prione~and~\pritwo~by comparing their abundances with theoretical model predictions of supernova (SN) yields from single non-rotating massive Population III stars in the range 9.6-100 M$_{\odot}$ of \citet{heger10}. 
We use only elements with available NLTE abundances because they are closest to {\em true} absolute abundances, which are required for a comparison with theoretical predictions. This exercise implicitly assumes that UMP stars 
are indeed {\em bona-fide} second-generation objects chemically enriched by a single Pop~III supernova, as commonly accepted in Galactic archaeology studies.

The model database includes 120 models and covers a range of explosion energies from 0.3 $\times$ 10$^{51}$ erg to 10 $\times$ 10$^{51}$ erg. For each energy, there are models with different (fixed) mixing amounts in the SN ejecta due to Rayleigh–Taylor instabilities \citep[see][for further details]{heger10}. We use their publicly available $\chi^2$ matching algorithm to determine which model fits our abundances best\footnote{Accessible at \url{http://starfit.org}}. We adopt a similar procedure as the one described in \citet{heger10}, using all available element measurements and upper limits up to atomic number Z = 30, as the nucleosynthetic origin of post iron-peak elements in Pop~III stars is quite uncertain.

The fitting algorithm treats Sc and Cu as model lower limits because they have multiple different nucleosynthetic pathways (e.g., there are additional production sites not included in the model data) and it ignores Li, Cr, and Zn by default. 
The same fitting procedure was also used in a number of other studies \citep[e.g.][]{placco15,gonzalez20}. 
 
 Models from \citet{heger10} reasonably reproduce chemical abundances in both cases. Fitting the abundance pattern of \prione~yields a progenitor mass of 14.4 M$_{\odot}$ with little mixing (f$_{\rm mix}$ = 3.98 $\times$ 10$^{-2}$), and a relatively low explosion energy of 1.2 $\times$10$^{51}$erg.
Similarly, the fitting algorithm returns a low progenitor mass (10.6 M$_{\odot}$) with mixing f$_{\rm mix}$ = 2.51 $\times$ 10$^{-1}$, and a low explosion energy (0.3 $\times$10$^{51}$ erg) for \pritwo~as well.

Besides the best-fit solution, the procedure also provides the ten best model fits ranked by their $\chi^2$ value.
If the best-fit solution is robust, we expect the $\chi^2$ value to rapidly increase between the first and second-best solutions \citep{placco15}.
For~\prione, the first two solutions have the same $\chi^2$ value, but the mass of the progenitor is the same with a slightly different explosion energy (1.2 $\times$10$^{51}$erg and 0.9 $\times$10$^{51}$erg in the first and second best-fit solutions, respectively). The progenitor mass returned by the third-best model is only slightly different (14.0 M$_{\odot}$) with a corresponding variation of the $\chi^2$ value of $\sim$ 18\%.
For~\pritwo~, the fitting algorithm returns the same progenitor mass and energies in the first six best-fit models.

According to \citet{placco15}, the quality of the match between the abundance pattern of a star to the yields is highly sensitive to the N abundance. In particular, the fitting algorithm predicts a very low N abundances ([N/H] < --6)  where only upper limits on nitrogen are determined. Moreover, carbon and nitrogen abundances are derived using the 1D models, thus they are significantly overestimated compared to those measured in 3D-NLTE model atmospheres \citep[e.g.][and references therein]{norris19}. To evaluate how changes in the C and N abundances reflect on the progenitor mass we estimated, we ran the matching procedure with both carbon and nitrogen reduced by 1 dex.  Overall, we obtain an improvement of the fit. 
Indeed, the best-fit model better reproduces carbon abundance, but there is no change in the progenitor model. We also determine the best fit parameters after excluding the N abundance from the observed pattern. We found no variations in the progenitor mass or explosion energy in both cases.

\begin{figure}

    \centering
       \includegraphics[width=0.9\columnwidth]{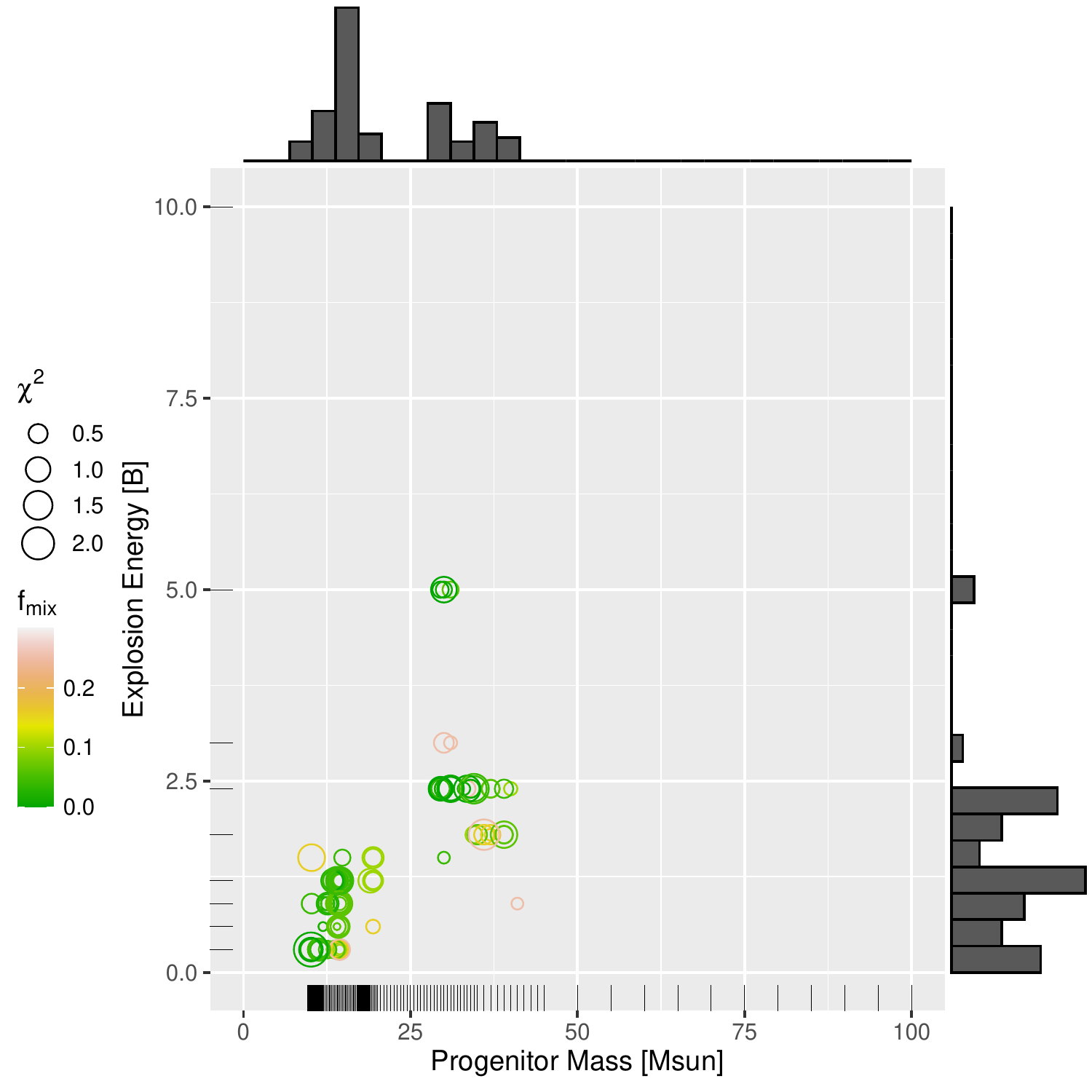}
       \includegraphics[width=0.9\columnwidth]{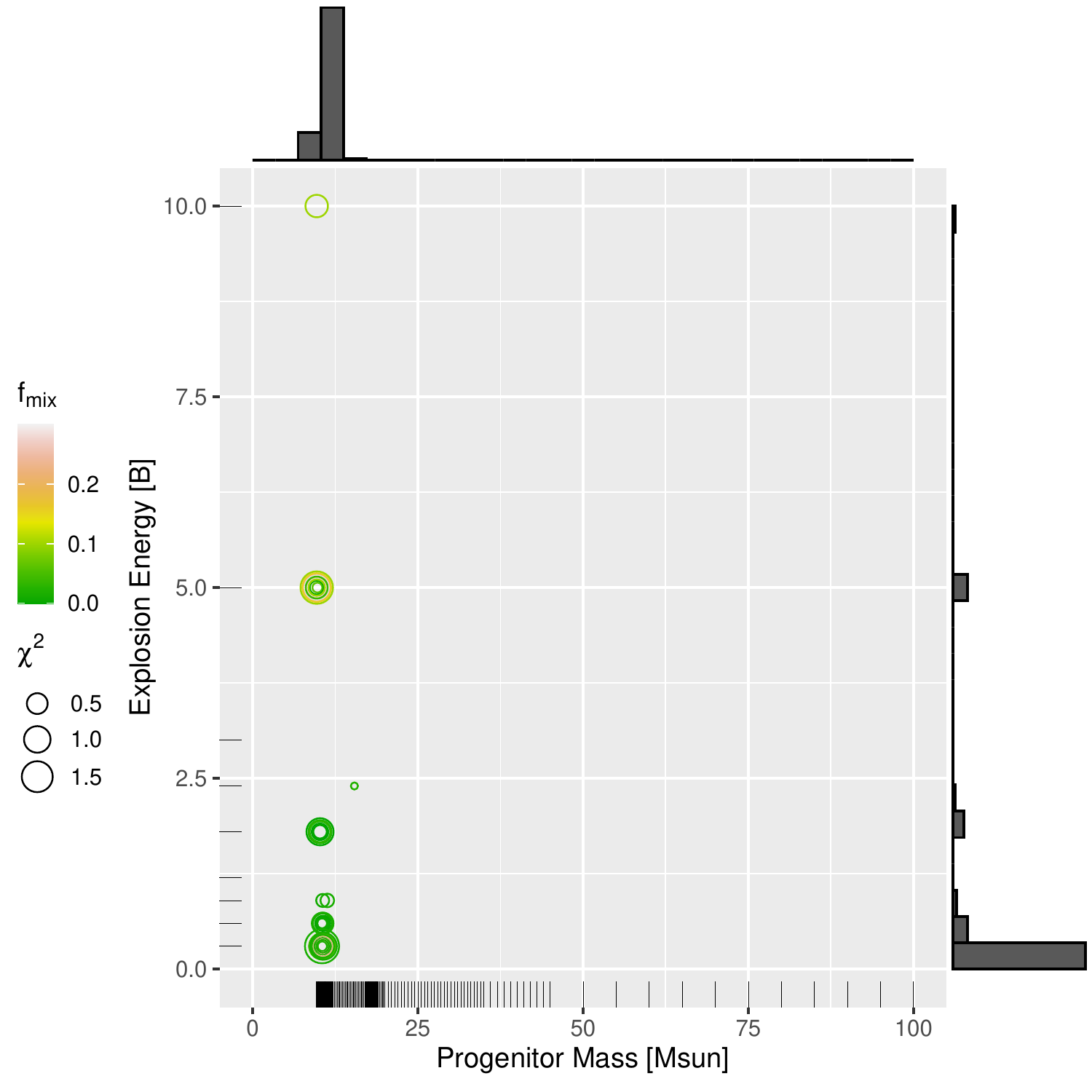}
            \caption{The run of the derived masses vs. explosion energies from 100 iterations of the $\chi^2$ matching algorithm using perturbed NLTE abundances are shown for ~\prione~and~\pritwo~in the top and bottom panel, respectively.
      Data points are colour-coded according to the amount of mixing characterising the best-fit model. The size of the data points is a function of the derived $\chi^2$ value. Marginal histograms for the progenitor mass and explosion energy are shown as well.
      The rug plots display the grid of masses and explosion energies from \citet{heger10}. There are 126 different masses in the range 9.6-100 M$_{\odot}$. Model stars differ by only 0.1-0.2 M$_{\odot}$ toward the lighter end of the mass range explored in \citet{heger10}  to take into account for their different pre-supernova properties \citep[see][]{heger10}.}
     
      \label{fig:montecarlo}

\end{figure}

To further quantify the uncertainty in progenitor masses computed by the $\chi^2$ matching algorithm and check the parameter degeneracy, we varied NLTE abundances for each star by their uncertainties and run the procedure using the perturbed abundances \citep[see][]{fraser17}. To this aim, we assumed that measurements were normally distributed and repeated this perturbation procedure 100 times for each star. The distribution of inferred progenitor masses as a function of the explosion energies, after fitting our perturbed abundance values are shown in Fig.~\ref{fig:montecarlo}. The figure also shows the resulting $\chi^{2}$ values and the amount of mixing of the different draws.

We find that supernova yields for models in a wide range of progenitor masses  between 10.10 and 41.00 M$_{\odot}$ well reproduce the chemical composition of ~\prione. The median value of the progenitor mass is 14.40 M$_{\odot}$, with a 1st quartile (Q1) of 14.00 M$_{\odot}$, a 3rd quartile (Q3) of 30.00 M$_{\odot}$. 
In addition to massive progenitors of 20-40 M$_{\odot}$ exploding in general with low energies (1-2.5 $\times$10$^{51}$ erg), we also find good fits for progenitors with masses $<$ 15 M$_{\odot}$ , with very low explosion energies ($<$1 $\times$10$^{51}$ erg). In particular, we computed a median value for the explosion energy of 1.2 $\times$10$^{51}$ erg (Q1= 0.9 $\times$10$^{51}$ erg and Q3=1.8 $\times$10$^{51}$ erg), with five notable outliers with explosion energy of 5 $\times$10$^{51}$ erg.  In general, models with less mixing are favoured, with a median value of f$_{\rm mix}$ =0.03981 (Q1=0.02512 and Q3=0.06310). The reduced mixing is likely due to the compact nature of the supernova progenitor star in this mass range \citep{heger10}.

\begin{figure}

    \centering
       \includegraphics[width=\columnwidth]{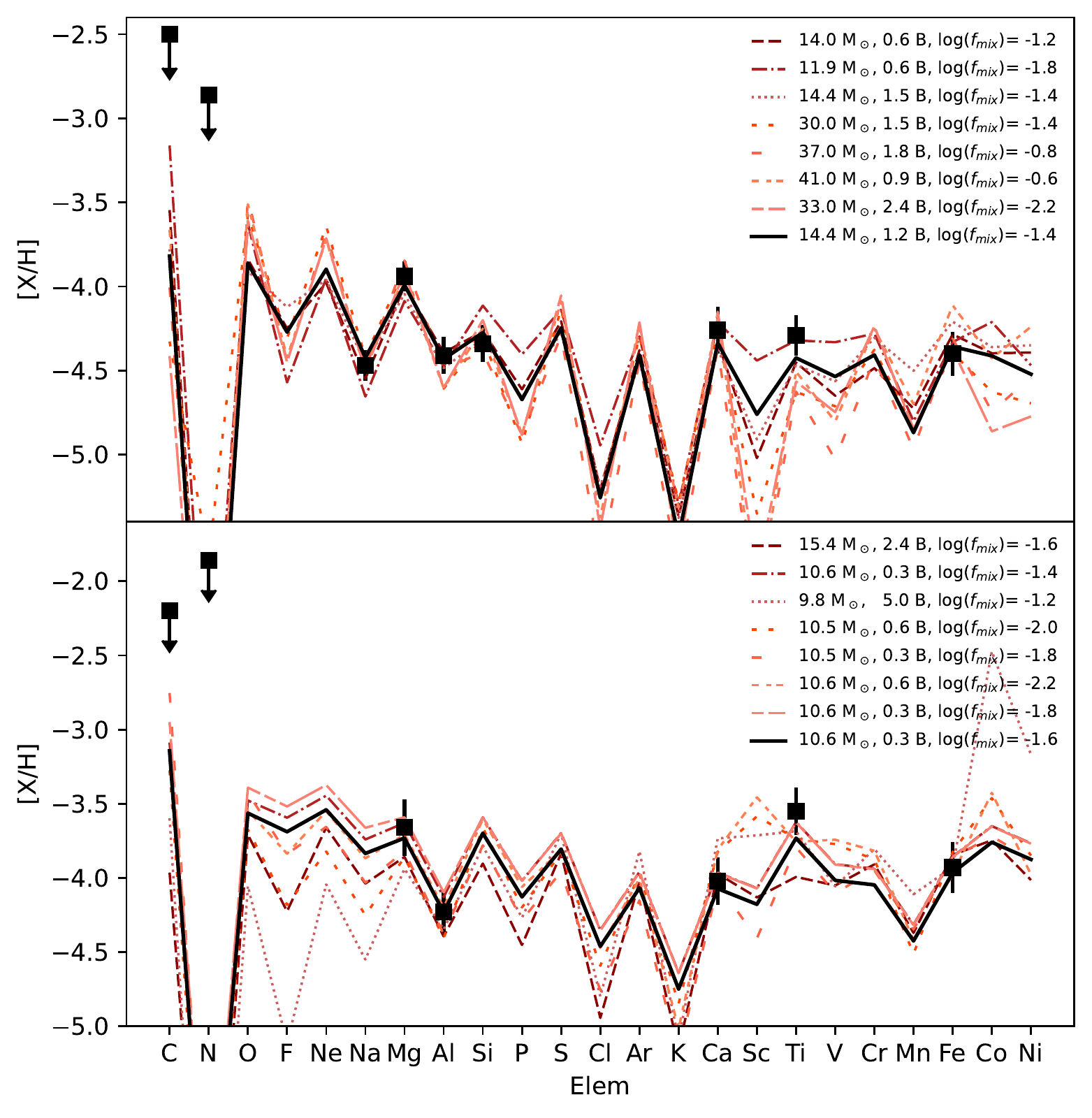}
        \caption{Supernova model yields assuming a selection of different progenitor mass, explosion energy, and mixing that best match the NLTE abundance patterns of ~\prione~and~\pritwo~(black squares) are shown in the top and bottom panel, respectively. Best fit model is plotted as a continuous black line.}
      \label{fig:progenitors}

\end{figure}

For~\pritwo, all the best-fit models matching its perturbed NLTE abundance pattern have low mass in the range between 9.70-15.40 M$_{\odot}$ (median value of 10.6 M$_{\odot}$, Q1= 10.50 M$_{\odot}$, Q3= 10.60 M$_{\odot}$). Also, they have generally experienced low mixing (median value of f$_{\rm mix}$ = 0.01585, Q1= 0.01585, Q3=0.08155) and low explosion energies (median value of 0.3 $\times$10$^{51}$ erg, Q1 = 0.3 $\times$10$^{51}$ erg, Q3 = 0.6 $\times$10$^{51}$ erg). Nine models have progenitor mass of 9.70-9.80 M$_{\odot}$ and unusually high explosion energy (>5 $\times$10$^{51}$ erg). Most probably, the lack of any Na constraint for ~\pritwo~causes the fit to always prefer lower-mass progenitors \citep{ishigaki18}.

We illustrate the yields from a selection of well-fitting models in Figure~\ref{fig:progenitors}. The models have been selected as those which best match observations for any explosion energy and mixing efficiency within the range of progenitor masses obtained from the 100 simulations for each star. Overall, both ~\prione~and~\pritwo~have abundance patterns have NLTE elemental abundance patterns that are well reproduced by the SN models of \citet{heger10}.

As discussed above, the existence of a nitrogen measurement is fairly important for the model fit procedure \citep[e.g.][]{placco15}. Indeed, in their study of 21 UMP stars observed at high-resolution, \citet{placco15} found that stars with upper limits on nitrogen have progenitor masses in the 10.6-15.0 M$_{\odot}$ range. In contrast, progenitor masses for stars with actual nitrogen measurements show a bimodal distribution toward higher masses (e.g., progenitor masses in the range 20.5–23.0 M$_{\odot}$ and 27.0–28.0 M$_{\odot}$). Thus, we caveat that the lack of nitrogen abundance measurements is possibly affecting the progenitor mass determination.

Finally, the measured abundance ratios for [Na/Mg], [Al/Mg], as well as [Ca/Mg] rule out any contamination from massive  (140 M$_{\odot}$ < M < 260 M$_{\odot}$ ) exploding as pair-instability supernovae \citep[PISNe;][]{heger02} for both stars. In particular, the PISN models predict low [(Na,Al)/Mg] and high [Ca/Mg] abundance ratios ([Na/Mg] = --1.5, [Al/Mg] = --1.20, and [Ca/Mg] $\sim$ 0.5-1.3; \citealp{takahashi18}) that are not observed in the analysed stars\footnote{We measure for~\prione~[Na/Mg] = -0.53, [Al/Mg] = -0.47,  and [Ca/Mg] = -0.32. The derived [Al/Mg] and [Ca/Mg] abundance ratios for~\pritwo are [Al/Mg] =-0.54  and [Ca/Mg] = -0.33.}.

.

\section{Conclusions}\label{conclusions}

We have presented a high-resolution spectroscopic 
follow-up of two UMP stars identified in the {\em Pristine} photometry and previously analysed in \citet{starkenburg18} (\prione) 
and \citet{aguado19} (\pritwo) based on the 
medium-resolution spectrum and in \citet{kielty20} (\pritwo) based on the high-resolution spectrum.
We now briefly summarise our findings.

\begin{itemize}
    \item We found that neither star shows any significant radial velocity variations during the observations in 2018 August (four nights; \citealp{starkenburg18})/ 2018 June (two nights; \citealp{kielty20}) - for \prione~and~\pritwo, respectively-- or those in 2020 (six nights; this paper), or between the two investigations. 
    \item We confirm the UMP nature of both stars, with measured Fe abundances of [Fe/H]$_{\rm LTE}$ = --4.79 $\pm$ 0.14 and [Fe/H]$_{\rm LTE}$ = --4.22 $\pm$ 0.12 for ~\prione~and~\pritwo, respectively. This translates to [Fe/H]$_{\rm NLTE}$ = --4.38 $\pm$ 0.13 and [Fe/H]$_{\rm NLTE}$ = --3.93 $\pm$ 0.12 when departures from LTE are taken into account.
    \item The upper limits we derive on the C abundances are compatible with the two stars having a moderate C overabundance. In particular, the upper limit and tentative measurement for ~\prione we derive in \S~\ref{const_on_carbon} using constraints from Al features lies on the border of the carbon-normal stars.
    We were able also to put an upper limit to the N abundance of both objects from the analysis of the NH feature at $\sim$3360~\AA.
   \item Measured abundances for Li, Na, Mg, Al, Si, Ca, Sc, Cr, Ti, and Ni
   in general exhibit good agreement with those of Galactic EMP and UMP stars. Li is depleted in ~\prione~probably because of its low effective temperature. On the other hand, the low Li abundance of \pritwo~is intrinsic as observed in many unevolved stars with [Fe/H]$<$--3.5 \citep{bonifacio18LI}. 
   We measure different [Sr/Fe] abundance ratios for ~\prione~and~\pritwo, with [\ion{Sr}{ii}/Fe]$_{\rm LTE}$ = --0.51 ([\ion{Sr}{ii}/Fe]$_{\rm NLTE}$ = --0.72) and [\ion{Sr}{ii}/Fe]$_{\rm LTE}$ = +0.06 ([\ion{Sr}{ii}/Fe]$_{\rm NLTE}$ = --0.05), respectively.
    The low [\ion{Sr}{ii}/Fe] abundance ratios together with the absence of any other heavier element features supports the interpretation of \citet{bonifacio15} that stars belonging to the low-carbon-band are  
    carbon enhanced metal-poor stars with no over-abundances of neutron-capture elements (CEMP-no according to the classification of \citealp{beers05}) as opposed to stars on the high-carbon-band in which the high carbon abundance is the consequence of mass-transfer in a binary system from companion in its asymptotic giant branch phase (e.g., \citealp{hansen16,yoon16}; but see also \citealp{arentsen19} for a discussion on the r\^ole of binarity in shaping the abundance pattern of CEMP-no stars).
    No detection of any RV variation further supports such scenario.

    \item Under the assumption that the chemical pattern of the two UMP stars directly reflect the composition of the parent gas cloud (e.g., stars have been enriched by a single Pop~III supernova),  their NLTE abundances suggest in both cases low-energy supernova progenitors with masses from 10.6 to 14.4 M$_{\odot}$ and energies in the range between 0.3-1.2 $\times$10$^{51}$ erg.

\end{itemize}

Ultra metal-poor stars are rare relics of the first stages of star formation and cosmic evolution \citep{beers05}.
As such, they guide our understanding on the first nucleosynthesis events and the first Pop~III stars \citep{bromm13},
the earliest phases of chemical enrichment \citep{frebel15}, as well as the formation of the first Pop~II stars with low-mass \citep{frebel07,klessen12,chiaki17}.

Obtaining reliable data for a significant sample of them is crucial to undertake large and systematic investigations of this important stellar population. Unfortunately,although the number of known UMP stars has greatly increased over the last years \citep[e.g.,][]{norris85,beers85,beers92,christlieb03,antony91,antony00, schuster96,allendeprieto00,keller07, starkenburg17,melendez16,schlaufma14,placco19},
the relatively small number currently available still precludes any full stellar population analysis. 
In the near future, on-going synergies between large photometric and spectroscopic surveys will revolutionise the field, yielding extremely large samples of stars with [Fe/H]$<$--3.0 and delivering additional UMP stars.

\section*{acknowledgements}
CL acknowledges funding from  Ministero dell'Università e della Ricerca through the Programme ``Rita Levi Montalcini'' (grant PGR18YRML1).
JIGH acknowledges financial support from the Spanish Ministry of Science and Innovation (MICINN) project AYA2017-86389-P, and also from the Spanish MICINN under 2013 Ram\'on y Cajal program RYC-2013-14875.
 NM, EC, PB, VH, and GK gratefully acknowledge support from the French National Research Agency (ANR) funded projects ``Pristine'' (ANR-18-CE31-0017). AA and NFM acknowledge funding from the European Research Council (ERC) under the European Unions Horizon 2020 research and innovation programme (grant agreement No. 834148). ES acknowledges funding through VIDI grant ``Pushing Galactic Archaeology to its limits'' (with project number VI.Vidi.193.093) which is funded by the Dutch Research Council (NWO).  The authors thank the International Space Science Institute, Berne, Switzerland for providing financial support and meeting facilities to the international team ``Pristine''. 
 
 This paper is based on observations collected at the European Organisation for Astronomical Research in the Southern Hemisphere and observations made with the Gran Telescopio de Canarias, operated by the GRANTECAN team at the Observatorio del Roque de los Muchachos, La Palma, Spain, of the Instituto de Astrof{\'\i}sica de Canarias.
This work has made use of data from the European Space Agency (ESA) mission
{\it Gaia} (\url{https://www.cosmos.esa.int/gaia}), processed by the {\it Gaia}
Data Processing and Analysis Consortium (DPAC,
\url{https://www.cosmos.esa.int/web/gaia/dpac/consortium}). Funding for the DPAC
has been provided by national institutions, in particular the institutions
participating in the {\it Gaia} Multilateral Agreement.

\section*{data availability}

The data underlying this article will be shared on reasonable request to the corresponding author.



\bibliographystyle{mnras}
\bibliography{biblio} 







\bsp	
\label{lastpage}
\end{document}